\documentclass[a4paper,12pt]{article}
\usepackage{times}
\usepackage{fullpage}
\usepackage{comment}
\usepackage[margin=1.9cm]{geometry}
\usepackage{graphicx}
\usepackage[utf8]{inputenc}
\usepackage{enumitem,amssymb}
\usepackage{courier}
\usepackage[dvipsnames]{xcolor}
\usepackage{bbding}
\usepackage{authblk}
\usepackage{pifont}
\usepackage[]{natbib}
\usepackage{eurosym} 
\usepackage[normalem]{ulem}
\usepackage{hyperref}
\hypersetup{
    colorlinks=true, %colorise les liens 
    breaklinks=true, %permet le retour à la ligne dans les liens trop longs 
    urlcolor= blue, %couleur des hyperliens 
    linkcolor= black,	%couleur des liens internes 
    citecolor=blue,	%couleur des références 
} 
\newcommand\apjl{ApJL}%
\newcommand\apjs{ApJS}%
\newcommand\aap{A\&A}%
\title{{\huge\bf SPHERE+ \\
\Large Imaging young Jupiters down  to the snowline }\\
\large White book submitted to ESO, Feb. 2020}
\author[1]{A. Boccaletti}           % lesia
\author[2]{G. Chauvin}              % ipag
\author[2]{D. Mouillet}             % ipag
\author[3]{O. Absil}                % liege
\author[4]{F. Allard}               % ens cral
\author[5]{S. Antoniucci}           % roma
\author[2]{J.-C. Augereau}          % ipag
\author[6]{P. Barge}                % lam
\author[7]{A. Baruffolo}            % padova
\author[8]{J.-L. Baudino}           % oxford
% ...
\affil[1]{LESIA, Observatoire de Paris, Universit{\'e} PSL, CNRS, Sorbonne Universit{\'e}, Univ. Paris Diderot, Sorbonne Paris Cit{\'e}, 5 place Jules Janssen, 92195 Meudon, France} 
\affil[2]{Univ. Grenoble Alpes, CNRS, IPAG, 38000 Grenoble, France}
\affil[3]{Space sciences, Technologies and Astrophysics Research (STAR) Institute, University of Liège, 19C allée du Six Août, 4000 Liège, Belgium}
\affil[4]{Univ Lyon, Univ Claude Bernard Lyon 1, Ens de Lyon, CNRS, Centre de Recherche Astrophysique de Lyon UMR5574, F-69230, Saint-Genis-Laval, France}
\affil[5]{INAF-Osservatorio Astronomico di Roma, via di Frascati 33, I-00078 Monte Porzio Catone, Italy}
\affil[6]{LAM (Laboratoire d’Astrophysique de Marseille) UMR 7326, Aix Marseille Univ., CNRS, CNES, Marseille, France} 
\affil[7]{INAF - Osservatorio Astronomico di Padova, Vicolo dell’ Osservatorio 5, 35122, Padova, Italy}
\affil[8]{Department of Astrophysics, Denys Wilkinson Building, Keble Road, Oxford, OX1 3RH, UK}
% ...

\date{}
\begin{document}
\maketitle

%%%%%%%%%%%%%%%%%%%%%%%%%%%%%%%%%%%%%%%%%%%%%%%%%%%%%%%%%%%%%%%%%%%%%%%%
%%%%%%%%%%%%%%%%%%%%%%%%%%%%%%%%%%%%%%%%%%%%%%%%%%%%%%%%%%%%%%%%%%%%%%%%
\pagebreak
\section{Executive summary}

The Spectro-Polarimetric High-contrast Exoplanet REsearch instrument \citep[SPHERE\footnote{\url{http://sphere.osug.fr}},][]{Beuzit2019} has now been in operation at the VLT for more than 5 years, demonstrating a high level of performance.  SPHERE has produced outstanding results using a variety of operating modes, primarily in the field of direct imaging of exoplanetary systems, focusing on exoplanets as point sources and circumstellar disks as extended objects.   The achievements obtained thus far with SPHERE ($\sim$200 refereed publications) in  different areas (exoplanets, disks, solar system, stellar physics...) have motivated a large consortium to propose an even more ambitious set of science cases, and its corresponding technical implementation in the form of an upgrade. The SPHERE+ project capitalizes on the expertise and lessons learned from SPHERE to push high contrast imaging performance to its limits on the VLT 8m-telescope. The scientific program of SPHERE+ described in this document will open a new and compelling scientific window for the upcoming decade in strong synergy with ground-based facilities (VLT/I, ELT, ALMA, and SKA) and space missions (\textit{Gaia}, \textit{JWST}, \textit{PLATO} and \textit{WFIRST}). While SPHERE has sampled the outer parts of planetary systems beyond a few tens of AU, SPHERE+ will dig into the inner regions around stars to reveal and characterize by mean of spectroscopy the giant planet population down to the snow line. Building on SPHERE's scientific heritage and resounding success, SPHERE+ will be a dedicated survey instrument which will strengthen the leadership of ESO and the European community in the very competitive field of direct imaging of exoplanetary systems. With enhanced capabilities, it will enable an even broader diversity of science cases including the study of the solar system, the birth and death of stars and the exploration of the inner regions of active galactic nuclei.\vspace{0.1cm}

Therefore, the main motivation for SPHERE+ relies on three key scientific requirements that are currently driving this project as well as our proposed instrumental concept. They can be summarized as follows (Sec. \ref{sec:sciencecases} for details): 

\begin{itemize}
    \item \texttt{sci.req.1} - {\bf Access the bulk of the young giant planet population down to the snow line} (3-10 au), in order to bridge the gap with complementary techniques (radial velocity, astrometry), taking advantage of the synergy with \textit{Gaia}, and to explore for the first time the complete demographics of young giant planets at all separations in order to constrain their formation and evolution mechanisms.
    
    \item  \texttt{sci.req.2} - {\bf Observe a large number of fainter (lower mass) stars} in the youngest ($1-10$\,Myr) associations (Lupus, Taurus, Chamaeleontis, Scorpius-Centaurus...), to directly study the formation of giant planets in their birth environment, building on the synergy with ALMA to characterize the architectures and properties of young protoplanetary disks, and how they relate to the population of planets observed around more evolved stars.
    
    \item  \texttt{sci.req.3} -  {\bf Improve the level of characterization of exoplanetary atmospheres} by increasing the spectral resolution in order to break degeneracies in giant planet atmosphere models and to measure abundances and other physical parameters, such as the radial and rotational velocities. Near-infrared will be the primary wavelength range utilized, but the visible range also delivers valuable information in the form of accretion tracers. 
\end{itemize}

Overall, the SPHERE+ top level requirements connected to the proposed science cases can be summarized by going {\bf closer}, {\bf deeper}, and {\bf fainter}. As we understand very well the limitations of SPHERE, the science requirements can be linked directly to the following instrumental requirements:
\begin{itemize}
    \item \texttt{tech.req.1} - Deeper/closer: {\bf increase the bandwidth of the xAO system} (typically 3kHz instead of 1kHz) and improve the correction of non-common path aberrations and coronagraphic rejection.
    \item \texttt{tech.req.2} - Fainter: include a {\bf more sensitive wavefront sensor} to gain 2-3 magnitudes for red stars.
    \item  \texttt{tech.req.3} - Enhanced characterization: {\bf develop spectroscopic facilities with significantly higher spectral resolution} compared to the current SPHERE Integral Field Spectrograph (IFS). In this respect, both medium ($R_\lambda=5\,000-10\,000$) and high ($R_\lambda=50\,000-100\,000$) resolutions are extremely valuable for the characterization of planetary atmospheres. 
\end{itemize}

We have devised an instrumental concept which minimizes the modifications of the current system as well as the associated risks and instrument downtime during the upgrade, while providing the potential for breakthrough scientific results.

SPHERE+ is supported by a large consortium with strong expertise in high contrast imaging, from theory, to data analysis and instrumentation. 

The project is timely, and given the international and very competitive context, it is mandatory to start the phase A as early as 2020. First of all, SPHERE+ will benefit from synergy with several facilities, in particular ALMA and \textit{Gaia}, to identify the best targets to image and characterize. Second, SPHERE+ will narrow down the sample of the most interesting systems deserving a thorough characterization, for instance with ERIS, Gravity/Gravity+, \textit{JWST}, and with first light instruments at the ELT (MICADO, HARMONI and METIS). Third, in the ELT framework for future instruments, the associated technical developments of SPHERE+ will directly inform and benefit development of PCS (Planetary Camera and Spectrograph), the most challenging project on the ground for direct imaging and characterization of the lightest planets (e.g. Super Earths).

%%%%%%%%%%%%%%%%%%%%%%%%%%%%%%%%%%%%%%%%%%%%%%%%%%%%%%%%%%%%%%%%%%%%%%%%
%%%%%%%%%%%%%%%%%%%%%%%%%%%%%%%%%%%%%%%%%%%%%%%%%%%%%%%%%%%%%%%%%%%%%%%%
\pagebreak

\section{Introduction}

\subsection{The Advent of Direct Imaging} 

In the 20 years following the first detection of exoplanets \citep{mayor1995}, exoplanetary science grew exponentially, revolutionizing our understanding of planetary system formation and evolution. The discovery of a wealth of planets, first by ground-based radial velocity and transit surveys then by the \textit{CoRoT} and \textit{Kepler} space telescopes, demonstrated the enormous diversity of exoplanets, yielding numerous planets (hot Jupiters, super-Earths, mini-Neptunes) which were notably different from the planets in our own solar system. In particular, the radial velocity surveys and the nominal \textit{Kepler} mission ($2009-2013$) revealed the demographics of the exoplanet population up to $\sim$ 1\,au, demonstrating that exoplanets (particularly the so-called super-Earths which are completely absent in our own solar system) are ubiquitous and a very common outcome of stellar system formation \citep{mayor2011,petigura2013}. At that time, these results significantly contradicted our classical core-accretion view of the planet formation process, suggesting that these planets might form further out in their planetary systems, then be subject to migration and atmospheric erosion. In parallel, long-term radial velocity surveys and exoplanet-triggered $\mu$-lensing events also suggested the presence of a reservoir of giant planets beyond 2-3\,au, concurring with the shorter-period demographic analyses. By probing the outer regions of the nearby extrasolar systems beyond 10\,au with sensitivity limits down to a few Jovian masses, SPHERE was thus expected to provide a complementary view of this demographic picture and provide a better understanding of how planetary systems form and evolve.

In parallel, the first generation of direct imaging instruments on both 10m-class telescopes and \textit{HST}, usually facility adaptive optics imagers with a coronagraphic add-on, enabled large systematic surveys of young, nearby stars. About 15 years after the first image of the prototypical $\beta$ Pictoris debris disk by \citet{smith1984}, these surveys revealed new structures and companions in the outer regions of extrasolar systems: large debris disks with strong asymmetries and gaps hinting at unseen planets orbiting in their cavities \citep[e.g., HD\,141569A and Fomalhaut,][]{augereau1999,kalas2005}, and the first planetary mass companions at large distances from their host stars (2M1207\,b, \citealt{2004A&A...425L..29C}, \href{https://www.eso.org/public/news/eso0515/}{ESO-PR0515}), likely formed by gravo-turbulent fragmentation or gravitational disk instabilities.  The implementation of differential techniques and advanced data processing algorithms in 2006-2007 finally led to the breakthrough discoveries of the first directly imaged exoplanets: the iconic $\beta$ Pic b \citep{Lagrange2009} with NaCo at the VLT, and the HR\,8799\,bcde planetary system \citep{Marois2010Nat} with the Keck and Gemini telescopes. Shortly before the implementation of SPHERE at the VLT in 2014, another exoplanet orbiting within 50~au from its host star, HD\,95086\,b, was discovered with NaCo \citep{Rameau2013}, extending the family portrait.

Given the large inner working angle and coarse resolution of \textit{HST} and the poor contrast limits of the first generation of ground-based imagers at the shortest angular separations, SPHERE was expected to unveil the inner regions between 10 and 100~au of young nearby stars for the first time, yielding a number of key results. First, with a systematic investigation of a large sample of young nearby stars, SPHERE would determine the demographics of planets in these regions down to a few Jupiter masses. Second, SPHERE would systematically characterize the atmospheric properties of already-imaged planets as well as newly imaged planets, transforming our understanding of the nature and structure of their atmospheres. Finally, SPHERE would investigate the morphology of the circumstellar disks in which such planets form and evolve, and study the interactions between these key ingredients of exoplanetary systems.

\subsection{SPHERE discoveries and achievements} 

The SPHERE instrument utilizes sophisticated instrumentation, in order to detect faint planetary signals at small angular separations from a bright star. A planet orbiting at 9\,au around a star at 30\,pc distance lies at a maximum angular separation of 300 mas, setting the typical angular resolution necessary for exoplanet imaging. As young planets are still warm, the star-planet contrast for a solar type star and a Jupiter-mass planet is $\sim 10^{-6}$, while for a mature (cold) Jupiter the contrast increases to 10$^{-8}$ at best (i.e. at the peak of its spectral energy distribution). For a super-Earth observed in reflected light, the planet-star contrast increases to 10$^{-9}$. Compared to the early discoveries of young and massive self-luminous giant planets like $\beta$ Pic b ($H$-band contrast of 10$^{-4}$ at 400 mas) with NaCo at the VLT, imaging of telluric planets with the ELT will require several orders of magnitude higher contrast at  10$\times$ smaller angular resolution. Innovative technological developments are therefore required to meet these ultimate goals. SPHERE brought a significant gain in performance compared to the first generation of planet imagers and now routinely achieves high-contrast performances down to 10$^{-5}-10^{-6}$ at typical separations of 300 mas. This ground-breaking performance was achieved due to decisions made more than 10 years to implement a 3-stage design combining: i/ high angular resolution using xAO, ii/ stellar light attenuation using coronagraphy, and iii/ speckle subtraction using differential imaging techniques (angular, spectral, and polarimetric). Powerful post-processing tools (iv/) contribute a final step, optimizing the stellar signal suppression. SPHERE's outstanding achieved contrasts and inner working angle provided the early motivation for systematic and large scale surveys to exploit this performance to search for exoplanets and disks around a large sample of young, nearby stars.

%---------
\begin{figure}[t]
\begin{center}
\includegraphics[height=5cm]{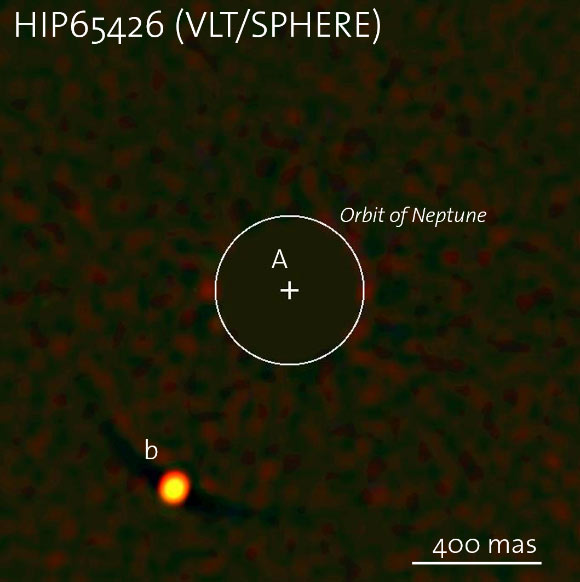}\hspace{0.5cm} 
\includegraphics[height=5cm]{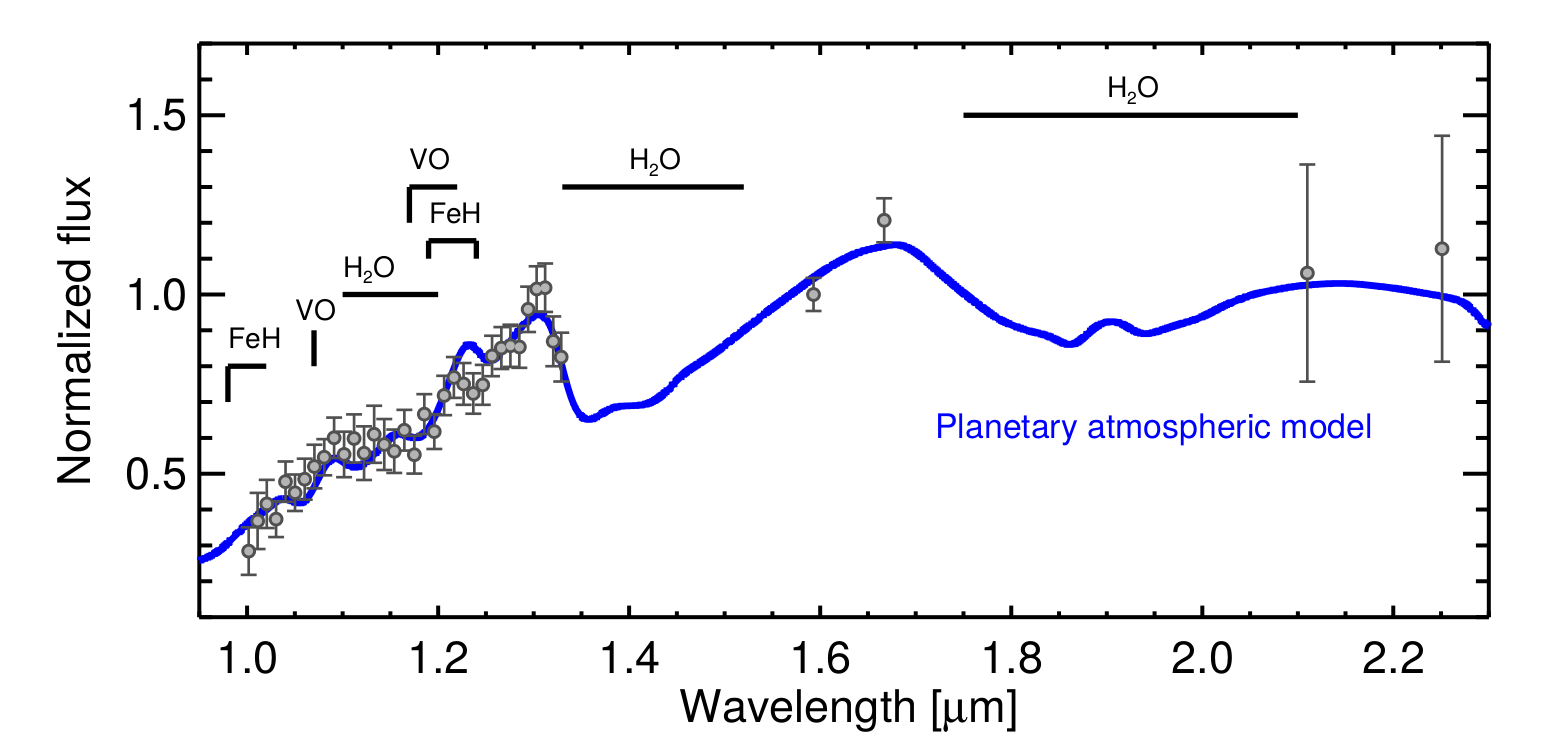} 
\caption{\textit{Left,} Composite IFS and IRDIS image of HIP 65426 A and b from
February, 2017. The first planet discovered with SPHERE  is well detected at a separation of $830\pm3$ mas and position angle of $150.0\pm0.3$ deg from HIP 65426.
 \textit{Right,} Near-infrared spectrum of HIP\,65426 b compared with the best-fit model atmosphere from the Exo-REM atmospheric model in blue ($T_{\rm{eff}}=1660\,\rm{K}$, $log(g)=4.5$, $f_{\rm{sed}}=1.5$, and $R=1.5\,R_{\rm{Jup}}$).}
\label{fig:hip65426}
\end{center}
\end{figure}

\subsubsection{Discovering and characterizing young, massive Jovian planets}

During the early phase of operation following SPHERE's first light in May 2014, deep characterization studies 
of several known young planetary systems highlighted SPHERE's detection capabilities, and both its astrometric and spectro-photometric performance.  These studies yielded new insight on a wide variety of companion objects, including: GJ\,758\,B, GJ\,504\,B and HD\,4113\,B \citep{Vigan2016,Bonnefoy2018,Cheetham2018a}, the coolest brown dwarf companions imaged to date, the young highly-eccentric brown dwarf companion around PZ Tel \citep{Maire2016}, the spectroscopic characterization of HD\,206893\,B,the reddest substellar companion discovered to date \citep{Delorme2017}, the newly discovered cool giant planet 51\,Eri\,b \citep{Samland2017}, the orbital and dynamical study of the young solar-system analogs architecture of HR\,8799 \citep{Zurlo2016, Bonnefoy2016} and the warm, dusty exoplanet HD\,95086\,b \citep{Chauvin2018}. These first studies paved the path towards the first SPHERE exoplanet discovery of a warm, dusty, and cloudy massive Jupiter around the young Lower-Centaurus Crux association member HIP\,65426 \citep[see Figure\,\ref{fig:hip65426},][\href{https://www.eso.org/public/announcements/ann17041/}{ESO-ANN17041}]{Chauvin2017}, which was followed by the first unambiguous discovery of a young planet recently formed within a transition disk and orbiting the young star PDS\,70 \citep[\href{https://www.eso.org/public/news/eso1821/}{ESO-PR1821}]{Keppler2018, Muller2018,mesa2019}. These latest breakthrough results demonstrate SPHERE's impressive capabilities and directly brought worldwide recognition in the exoplanetary research field. After 5 years of operation, the key SPHERE achievements in this domain can be summarized as follows:
%---------
\begin{figure}[t]
\begin{minipage}[c]{8.5cm}
\includegraphics[width=8.cm]{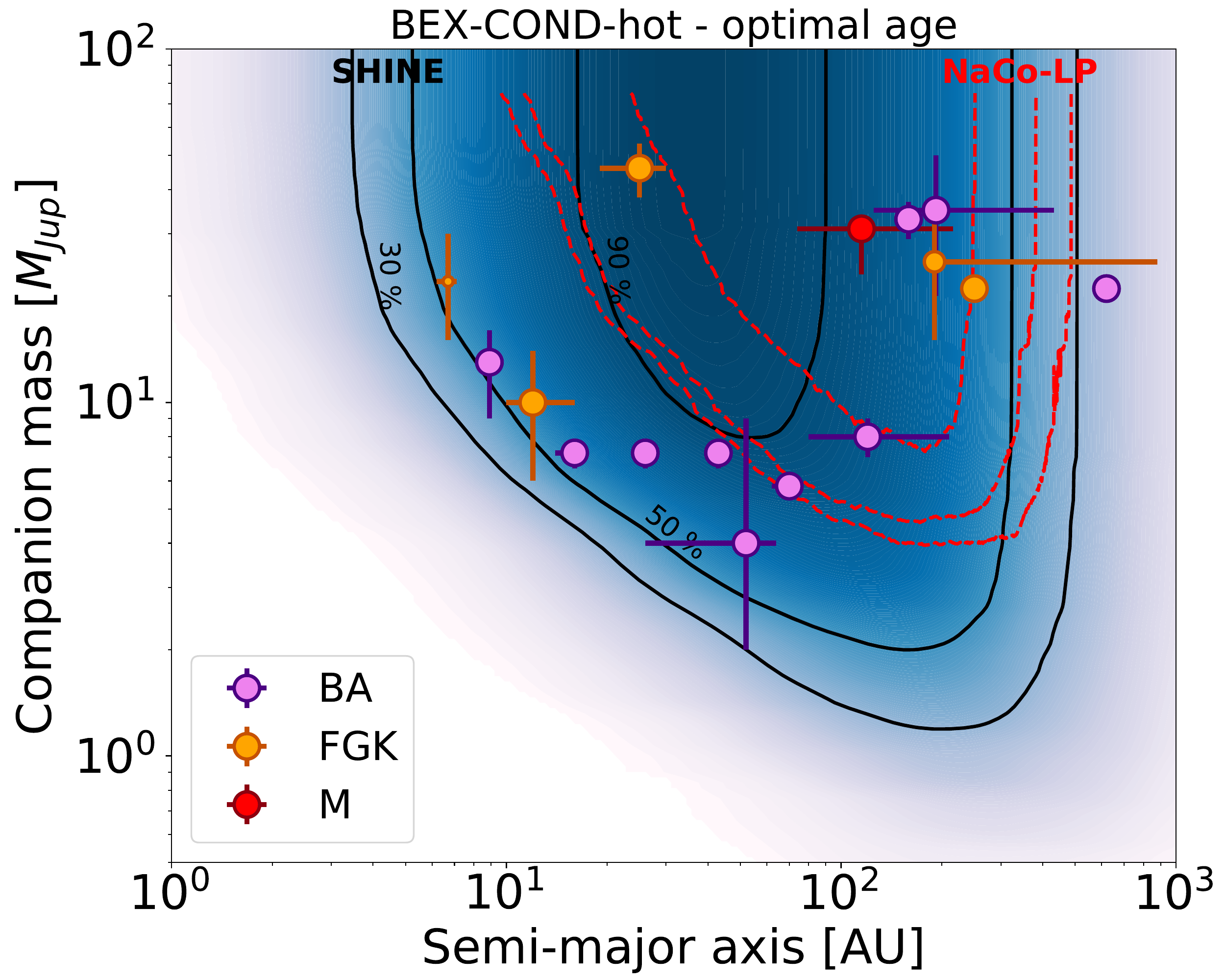} 
\end{minipage}
\begin{minipage}[c]{8.cm}
\caption{ SHINE survey completeness for a first sample of 150 young, nearby stars. The plot gives the numbers of stars around which the survey is complete for sub-stellar companions as a function of mass and semi-major axis. The mass conversion of the detection limits is based on the optimal stellar ages and on the BEX-COND-hot evolutionary models \citep{Marleau2019}. The colored circles represent the detected sub-stellar companions in the sample, with the color indicating the spectral type of the primary star (BA, FGK, or M). The NaCo-LP survey completeness is reported in \textit{red} for comparison. }
\label{fig:shinestat}
\end{minipage}
\end{figure}
%---------
\begin{itemize}

    \item Physics of exoplanets: in addition to the discovery of new exoplanets, SPHERE acquired unprecedented near-infrared photometry and spectra of young giant planets, enabling the exploration of the physical processes at play in their atmospheres (see Figure\,\ref{fig:hip65426}).
    The early characterization of the HR 8799 d and e planets by \citet{Zurlo2016} and \citet{Bonnefoy2016} demonstrated that the peculiar properties of young L/T type planets due to the low-surface gravity conditions in their atmospheres could successfully reproduce the spectral energy distributions of these planets.  In particular, low surface gravity leads to an enhanced production of thick, dusty clouds composed of sub-$\mu$m grains made of corundum, iron, enstatite, or forsterite, generating the observed red colors of these planets.
    Similar trends were found for the young planets 51\,Eri\,b, HIP\,65426\,b, HD 95086\,b, and PDS 70\,b and c \citep{Samland2017, cheetham2019, Chauvin2018, Muller2018, mesa2019}. These results were complemented by spectral characterization of young brown dwarf companions, such as HR\,3549\,B, HR\,2562\,B, and HIP\,64892\,B \citep{mesa2016,mesa2018,cheetham2018b}. Enabled by SPHERE's spectro-photometric capabilities, we are now building a detailed spectral sequence of young brown dwarfs and exoplanets extending from late-M to T-types that allows us to explore the effect of effective temperature, surface gravity (pressure), composition, clouds, and thermodynamical instabilities in the atmospheres of substellar objects down to young Jupiter analogs.   

    \item Planetary architectures: At young ($\lesssim$ 10\,Myr) ages, the presence of cavities, spirals, and kinematic perturbations in protostellar disks are likely to be caused by protoplanets in formation. In HD\,169142, we identified several structures in Keplerian motion, blobs, spiral arms, and one potential planet in formation \citep{2019A&A...623A.140G}. For systems older than 10 Myr, gas will have dispersed, giant planet formation will be complete, and the dynamics of planetesimals will be influenced by the presence of these giant planets.  SPHERE's high contrast performance and astrometric precision at level of 1-2 mas enables accurate monitoring of exoplanet orbits over time.  This offers the potential to study not only exoplanet orbits, but also the global system architecture including planet-planet and planet-disk interactions and the system stability. The $\beta$ Pictoris system provides the archetypical case for this sort of detailed study of system architecture.  The orbital monitoring of $\beta$ Pic\,b - combining NaCo and SPHERE - enabled continuous monitoring of the planet’s orbit around its star, passing behind and in front of  the star. The planet was observed down to 130\,mas in projected separation in Fall 2016 and recovered at 150\,mas in Fall 2018, after its passage in front of the star. To our knowledge, no other instrument could reach such a high contrast capability \citep[][
\href{https://www.eso.org/public/videos/esocast183a/}{ESOcast-183}]{Lagrange2019}. This accurate orbital monitoring campaign has confirmed that the planet is shaping the inner warp of the circumstellar disk and is responsible for the exocometary activity known for decades.  Similar studies have been conducted for the young solar analogs HR 8799 and HD 95086 \citep{Zurlo2016, Chauvin2018} as well as for 51 Eri and HR\,2562 \citep{maire_2019,2018A&A...615A.177M}. Here again, the planet physical and orbital properties can be compared to the dust spatial distribution and architecture, and reveal configurations very similar to that of our own solar system (with the asteroid and the Kuiper belts separated by the existence of giant planets). These systems provide precious laboratories to refine our understanding of the formation of our own solar system. 

    \item Occurrence \& formation: With a significant number of new exoplanet discoveries since the 51 Peg b announcement, theories of planetary formation have drastically evolved to digest these new observational constraints. However, we are still missing the full picture and there are a number of key holes in our understanding, such as the physics of accretion to form planetary atmospheres, the formation and evolution mechanisms to explain the existence of giant planets at wide orbits, the physical properties (including the mass-luminosity relation) of young Jupiters, and the influence of the stellar mass and stellar environment in the planetary formation processes. Neither core accretion plus gas capture nor disk fragmentation driven by gravitational instabilities can globally explain all current observables from the current census of exoplanets.   Moreover, we are still lacking a global view of the occurrence of exoplanets at all orbital periods. In this context, the SpHere INfrared survey for Exoplanets (SHINE) of 600 stars from the SPHERE Guaranteed Time Observations (GTO) provides unique statistical constraints on the demographics of giant planets beyond 10\,au (see the mean survey detection probability in Figure~\ref{fig:shinestat} compared to that from the first generation of planet imagers). Given the unprecedented detection performances achieved by SPHERE, the early statistical results demonstrate that giant planets are relatively rare beyond 10\,au, finding a frequency of $6\pm3$\,\% for planetary systems hosting at least one massive ($\ge2$\,M$_{\rm Jup}$) giant planet between 10 to 500 au (to be consolidated at the end of the survey). The size and stellar properties of the SHINE sample will still enable constraints on the variation of exoplanet occurrence rate with stellar mass and age. The final survey completeness will also be directly compared to the current predictions from planetary formation models to test the efficiency of each physical process (core accretion, gravitational instability, migration, dynamical scattering, ejection, see \citealt{Vigan2017}.  However, the current performance of SPHERE will not allow us to fully bridge the gap in planet separation in order to image planets detected via complementary techniques (transit, radial velocity, $\mu$-lensing and astrometry).  Thus, further improvements in contrast and especially inner working angle will be necessary to offer a global vision of planet occurrence at all separations.
\end{itemize}

\subsubsection{Dissecting planet forming disks}

%---------
\begin{figure*}[t]
\begin{center}
\includegraphics[height=5.cm]{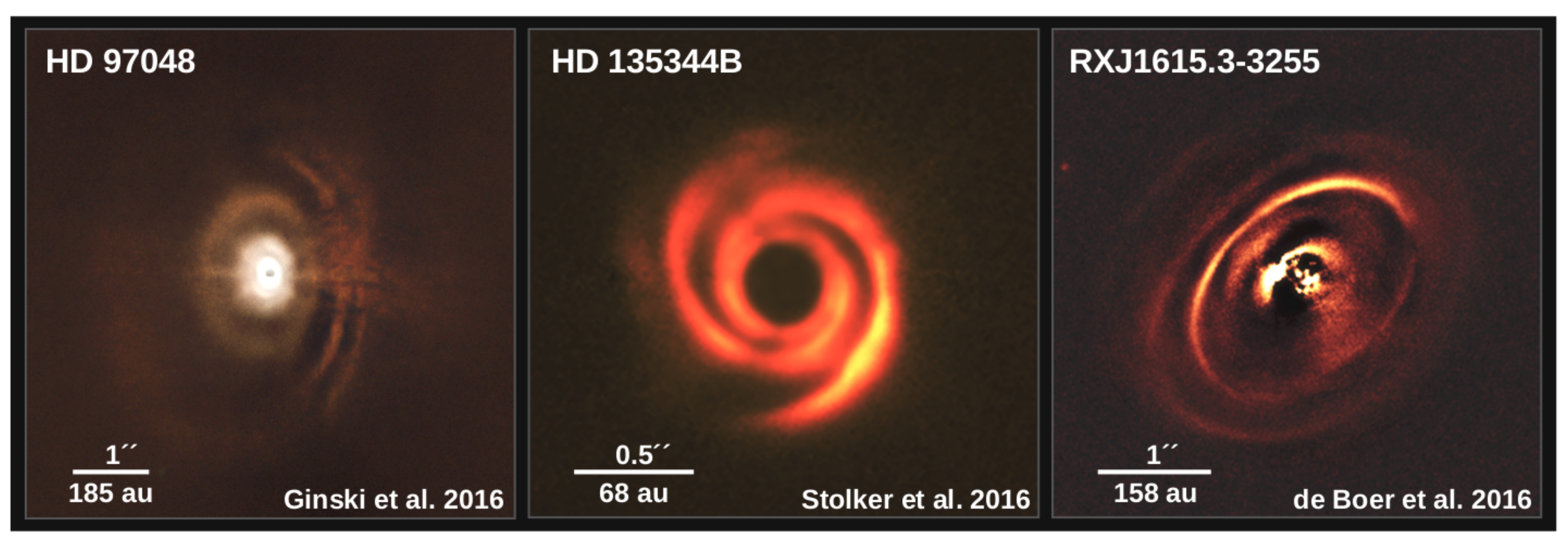}
\caption{Sharp SPHERE observations revealing striking features (rings, gaps, shadows, spirals) in planet-forming disks around the young stars HD\,97048, HD\,135344\,B, and RXJ\,1615.}
\label{fig:disks}
\end{center}
\end{figure*}
\begin{figure*}[!t]
\begin{minipage}[c]{10.5cm}
\includegraphics[width=10cm]{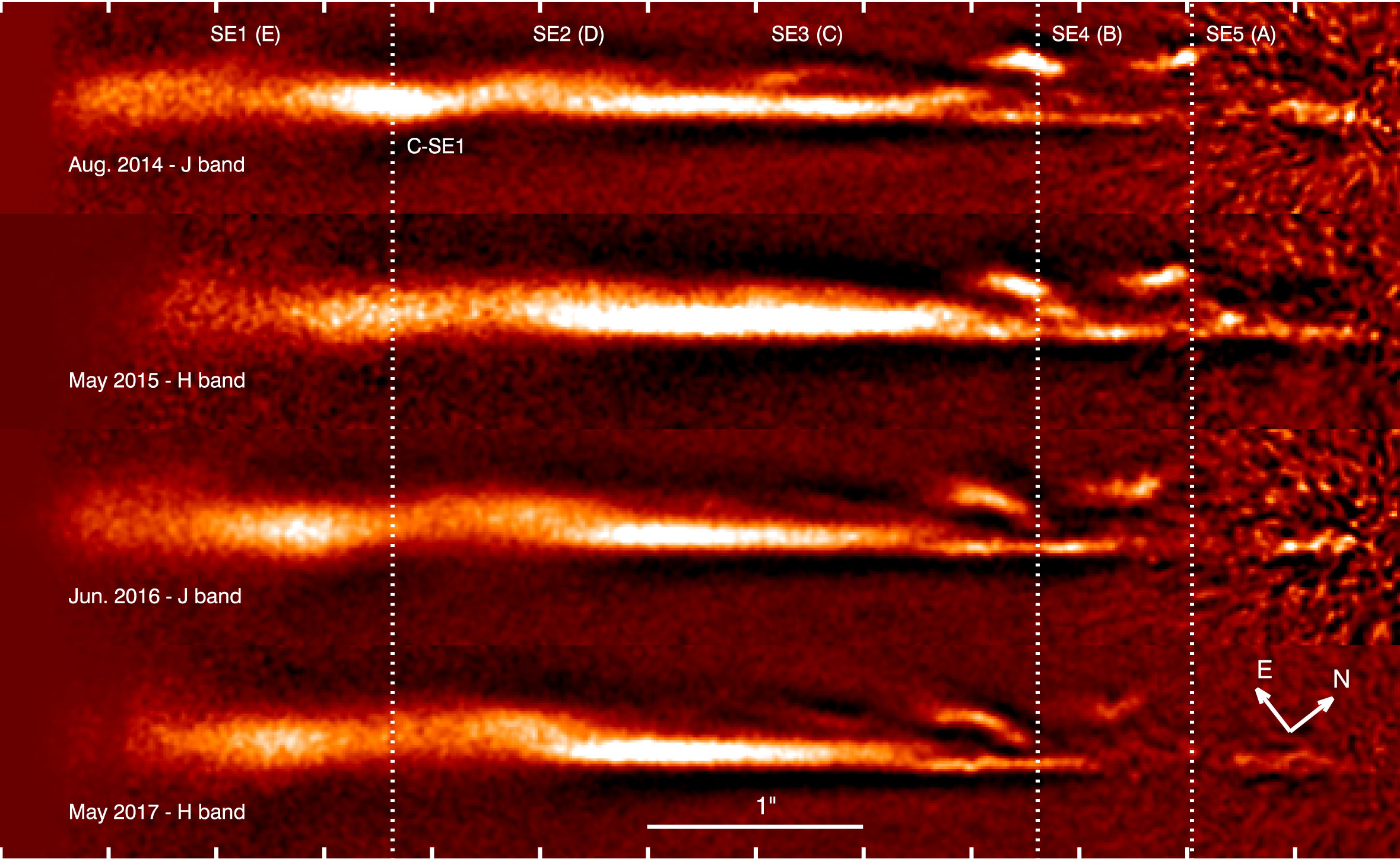}
\end{minipage}
\begin{minipage}[c]{6.5cm}
\caption{Fast moving structures observed in the debris disk of AU\,Mic, from Aug. 2014 to May 2017, and possibly triggered by the action of the stellar wind on an undetected planet or a reservoir of dust
\citep{Boccaletti2018}.}
\label{fig:disksaumic}
\end{minipage}
\end{figure*}
%---------

One of the greatest achievements of SPHERE has been to reveal the fine structures within imaged circumstellar disks, yielding unprecented information regarding the structure of planetary systems during or right after the planetary formation phase (see Figure\,\ref{fig:disks} and Figure\,\ref{fig:disksaumic}). SPHERE has observed two types of disks, protoplanetary disks ($\lesssim10$\,Myr) where both gas and dust are present in which we expect planets are still forming, and debris disks ($\gtrsim10$\,Myr), which are dominated by second generation dust (from collisions between planetesimals) and where planets have already formed. The transformative capabilities of SPHERE in the context of circumstellar disks relies on two aspects of its design: 1) SPHERE has a sufficiently large field of view to capture a full picture of disk structure and 2) SPHERE provides two main observing modes allowing total intensity and polarized intensity imaging both in the visible and near IR. With these capabilities, we were able to resolve new structures in known disks, to obtain the first images of disks suspected from their IR excess, and to serendipitously discover new disks (see below). 

SPHERE has spatially resolved many previously unseen disk structures, and in some cases characterized them either spectroscopically or polarimetrically.   Among the most notable results, SPHERE has discovered: large spiral arms \citep[MWC\,758,][]{Benisty2015}, multiple rings or gaps 
\citep[TW\,Hya, RXJ\,1615, HD\,97048,][]{vanBoekel2017, deBoer2016,  Ginski2016}, variable shadows \citep[HD\,135344\,B, V4046 Sgr,][]{Stolker2016, 2019NatAs...3..167D}, and central cavities 
\citep[HD\,169142, PDS\,70,][]{Pohl2017, Ligi2018, Keppler2018} in protoplanetary disks. In debris disks, SPHERE has also resolved several sharp Kuiper-like belts 
\citep[HD\,106906, HR\,4796, HD\,114082, GSC 07396-00759,][]{Lagrange2016, Milli2017, Wahhaj2016, Sissa2018}, multiple belts in gas-rich debris disks \citep[HD\,61005,  HD\,141569A, HIP\,73145, HIP\,67497,][]{Olofsson2016, Perrot2016, Feldt2017, Bonnefoy2017} and moving structures \citep[AU\,Mic,][]{Boccaletti2015, Boccaletti2018}. The vast majority of these structures could be explained by the presence of planets 
\citep{Dong2015a, Dong2015b, LeeChiang2016, Sezestre2017}
or by massive collisions \citep{Jackson2014, Kral2015}. Clearly, disk science is closely connected to exoplanet science, as reinforced by the results described above and especially the case of PDS\,70\,b, where a planet was discovered right inside the cavity of a gas-rich disk. Spectral and polarimetric information collected by SPHERE has also played an important role in putting constraints on dust properties (size distribution for instance or mineralogy). In only 5 years, the contribution of SPHERE to disk science is already impressive with a total of 50 disks resolved for the first time in scattered light, and about 60 related papers published, strengthened by a very strong synergy with ALMA. In fact, SPHERE has been able to image for the first time 11 debris disks (out of a total of 45 directly imaged debris , i.e. 25\%) and 39 protoplanetary disks (out of the 200 known in nearby star forming regions, i.e. 20\%), hence competing (in only 5 years!) with \textit{HST} which achieves a better sensitivity at large separations, but poorer angular resolution.
 
\begin{figure*}[t]
\begin{minipage}[c]{11cm}
\includegraphics[width=10.5cm]{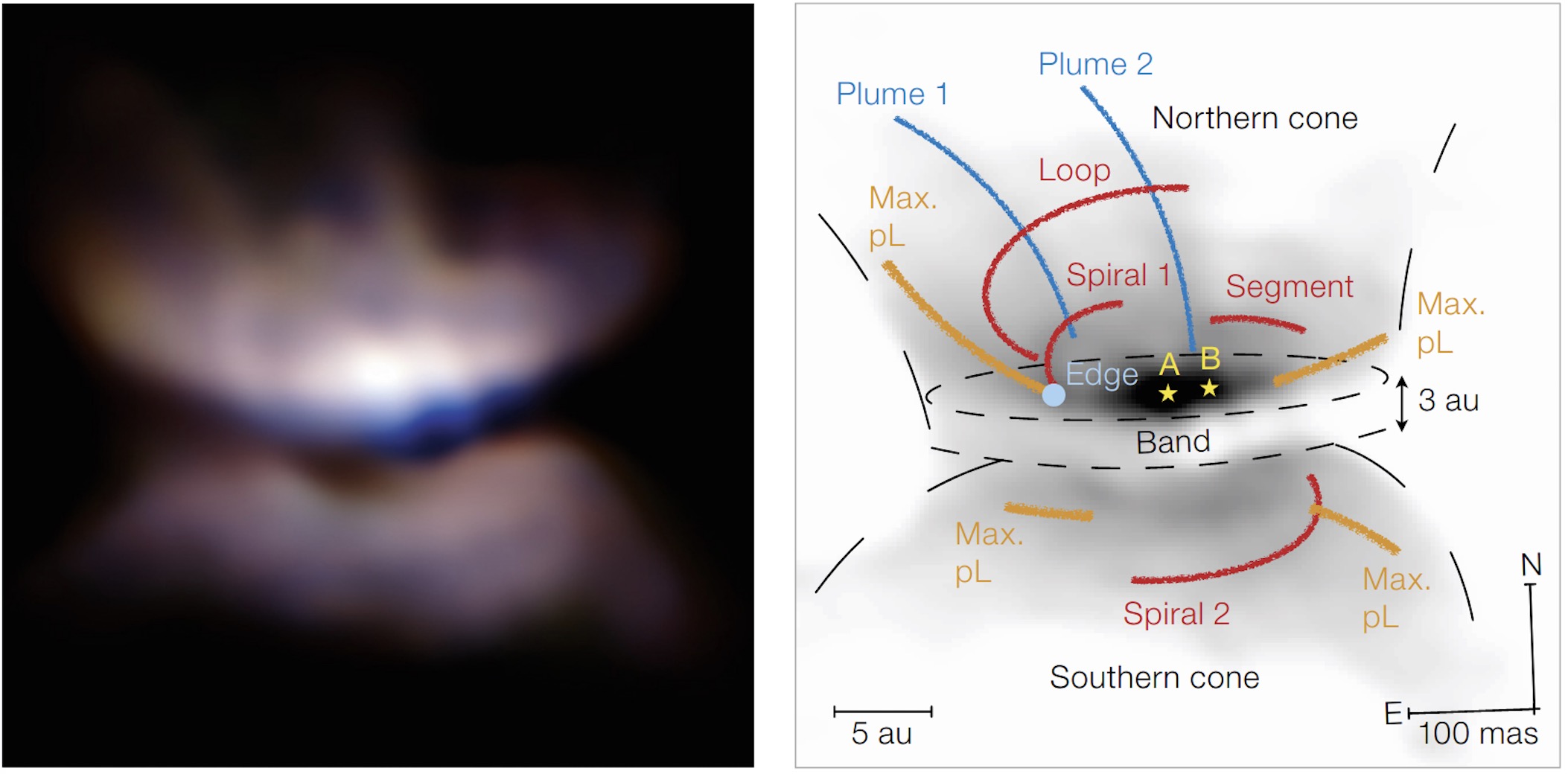}
\end{minipage}
\begin{minipage}[c]{6cm}
\caption{\textit{Left,} Colour composite intensity image of L2 Pup assembled from SPHERE/ZIMPOL V- and R-band images. \textit{Right,} nomenclature of the observed structures. 
}
\label{fig:L2pup}
\end{minipage}
\end{figure*}

\subsubsection{Galactic, stellar, and solar system physics}% (> Eric)
 
Although SPHERE was designed specifically to image planets and disks, its sophisticated instrumentation has led to breakthrough discoveries in a broad variety of astrophysical topics including galactic, stellar, and solar system physics. For instance, SPHERE obtained revolutionary results for the study of the late stages of stellar evolution which are characterized by a high mass loss rate, leading to the chemical enrichment of the Galaxy and to the final end product of the star's evolution. This mass loss is poorly understood, and results from the combination of convection, pulsation, dust and molecule formation, and radiation pressure. The resolution achieved with ZIMPOL (20\,mas), combined with its polarimetric capabilities (enabling the detection of dusty clumps) were the key to directly imaging the surface of a few giant stars, such as Betelgeuse \citep{Kervella2016} and R Doradus \citep{Khouri2016}. These studies also mapped gas next to the surface of stars and showed that the molecular opacity was time-dependent, and due to the formation of molecules above convective cells. The formation of these molecules and shocks due to a combination of pulsation and convection can lead to dust formation, as observed for W Hya \citep{Ohnaka2017} and Mira Ceti \citep{Khouri2018}. For oxygen-rich stars, SPHERE showed for the first time that the silicate grains around evolved stars are large (micron-sized), and that their mass loss can be explained by scattering by large grains, as observed for VY CMa \citep[][\href{https://www.eso.org/public/news/eso1546/}{ESO-PR1546}]{scicluna2015} and W Hya \citep{Ohnaka2016}. This led to a significantly better understanding of the impact of companions on the mass loss from these evolved stars, as most stars are born in multiple systems. The role of binary interactions on mass loss was further illustrated by the unprecedentedly detailed mapping of the dusty spiral formed by the interaction of a archetypical Wolf Rayet star with its O star companion in WR 104 \citep{soulain2018} and observations of the prototypical symbiotic system (an interacting binary system with an Asymptotic Giant Branch (AGB) star and a compact companion) R Aquarii, which resolved, for the first time,its binary companion and a precessing jet due to their interlace interaction \citep[][\href{https://www.eso.org/public/news/eso1840/}{ESO-PR1840}]{schmid2017}. Finally, for the nearby AGB star L2 Puppis (see Figure \,\ref{fig:L2pup}), SPHERE resolved a close companion and clear signs of interactions, forming an equatorial disk with ejection of plumes perpendicular to the disk \citep[][\href{https://www.eso.org/public/news/eso1523/}{ESO-PR1523}]{kervella2015}. This is the first direct image of the precursor of the commonly-observed and spectacular butterfly planetary nebulae shape and certainly represents the future of our own solar system. SPHERE also provided the first direct insights on the launching region of jets from young stars, which is crucial for determining their feedback on the disk and requires probing distances within a few au from the driving source (tens of mas for close-by Young Stellar Objects). In particular, \citet{antoniucci2016} observed the jets from Z CMa, deriving the first direct mass loss rate estimate for a FUor object and detecting a wiggling possibly indicative of an unseen close-in companion, while \citet{garufi_2019} analyzed the jet from RY Tau and its interaction with the complex circumstellar structures of the source. 

\begin{figure*}[t]
\begin{minipage}[c]{10cm}
\includegraphics[width=9.cm]{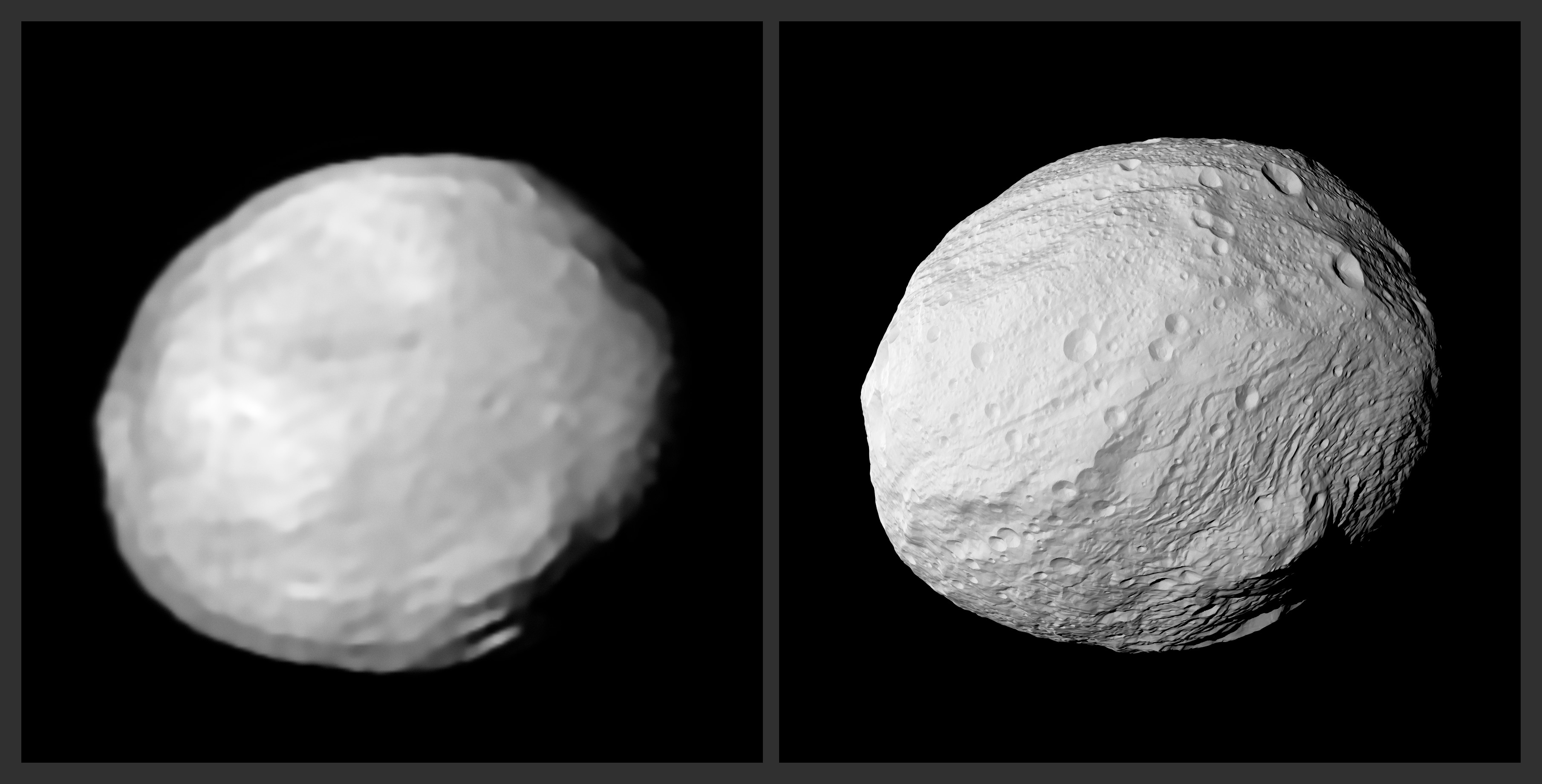}
\end{minipage}
\begin{minipage}[c]{7.cm}
\caption{ZIMPOL’s impressive image of the asteroid Vesta 
shown together with a synthetic view derived from \textit{NASA} space-based data. ZIMPOL image shows Vesta’s main features: the giant impact basin at Vesta's south pole, and the mountain at the bottom right \citep{Vernazza2018}. }
\label{fig:asteroids}
\end{minipage}
\end{figure*}

\begin{figure*}[t]
\begin{minipage}[c]{10cm}
\includegraphics[width=9cm]{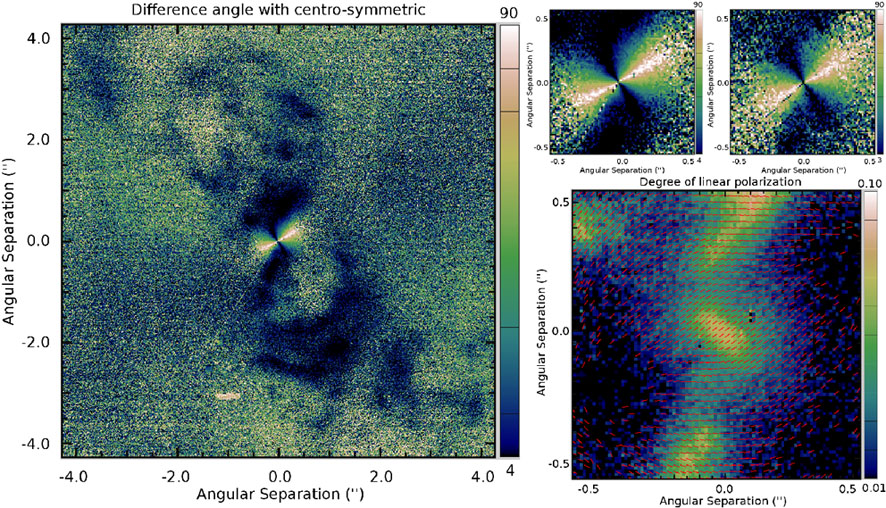}
\end{minipage}
\begin{minipage}[c]{7cm}
\caption{Polarimetric observations of NGC\,1068, showing the deviation of the polarization angle map with respect to a purely centro-symmetric pattern, and presumably tracing the suspected dust torus \citet{Gratadour2015}.}
\label{fig:agns}
\end{minipage}
\end{figure*}

The unique and versatile capabilities of SPHERE also opened new doors in ground-based asteroid exploration, namely, geophysics and geology to obtain resolved images of the surfaces of asteroids. Making use of the rotation of these objects and 3D modeling, the accurate shapes (and volumes) of asteroids can be determined. For asteroids in binary systems, this can lead to an accurate mass (and thus density) determination. Comparisons with results from on site space missions demonstrate that ZIMPOL can reveal with great accuracy the topography and albedo of asteroids (see Figure\,\ref{fig:asteroids}, e.g. the Vesta observations by \citealt{2019A&A...623A...6F}).  SPHERE has resolved and determined the 3D shapes of several asteroids (Elektra, \citealt{2017A&A...599A..36H}; Julia, \citealt{2018A&A...618A.154V}; Daphne, \citealt{2019A&A...623A.132C}; Hebe, \citealt{2017A&A...604A..64M}; (3) Juno, \citealt{2015A&A...581L...3V}; (16) Psyche, \citealt{2018A&A...619L...3V}; see \href{https://www.eso.org/public/news/eso1910/}{ESO-PR1910} and 5 pictures of the week for Ceres, \href{https://www.eso.org/public/images/potw1536a/}{ESO-POTW1536}; Asteroids, \href{https://www.eso.org/public/images/potw1725a/}{ESO-POTW1725} and \href{https://www.eso.org/public/images/potw1617a/}{1617}; Titan, \href{https://www.eso.org/public/news/eso1417/}{ESO-PR1417}; Vesta, \href{https://www.eso.org/public/images/potw1826a/}{ESO-POTW1826}). Studying the density and surface craters of these asteroids enabled the first ground-based study of the geology of these asteroids, improving our understanding of their formation and their parent bodies and families.
 
Finally, the angular resolution and sensitivity reached by SPHERE has improved our knowledge of stellar clusters and AGNs. \cite{Khorrami2016} studied the core of the young massive cluster NGC 3603 and showed that its central mass function was different than previously assumed, as previous studies were limited by crowding and the lack of dynamic range. Observation of the extragalactic cluster R136 found that the stars thought to be the most massive in the Universe had optical companions, and thus their masses had been overestimated \citep{2017A&A...602A..56K}. Finally, the circumnuclear optically thick material hiding the central core of the Active Galactic Nuclei NGC 1068 (Figure \ref{fig:agns}) was mapped with SPHERE. These data reveal an extended nuclear torus at the center of NGC 1068 \citep{Gratadour2015}, consisting of a structured hourglass-shaped bicone and a compact elongated (20\,pc $\times$ 60\,pc) nuclear structure perpendicular to the bicone axis.  SPHERE is thus a very versatile instrument, able to tackle a large variety of fundamental astrophysical questions including various ones that were not anticipated, such as the first determination of the mass of Proxima Cen from a microlensing event \citep{zurlo2018}.

%%%%%%%%%%%%%%%%%%%%%%%%%%%%%%%%%%%%%%%%%%%%%%%%%%%%%%%%%%%%%%%%%%%%%%%%
%%%%%%%%%%%%%%%%%%%%%%%%%%%%%%%%%%%%%%%%%%%%%%%%%%%%%%%%%%%%%%%%%%%%%%%%
%\pagebreak

\section{SPHERE+ Science cases}
\label{sec:sciencecases}

\subsection{Exoplanets}
\label{sec:exopla}

%%%%%%%%%%%%%%%%%%%%%%%%%%%%%%%%%%%%%%%%%%%%%%%%%%%%%%%%%%%%%%%%%%%%%%%%%%%%%
\subsubsection[Demographics of giant planets: bridging the gap down to the snow line]{Demographics of giant planets: bridging the gap down to the snow line (\texttt{sci.req.1)}}

\begin{figure*}[t]
\begin{center}
\includegraphics[height=5.5cm]{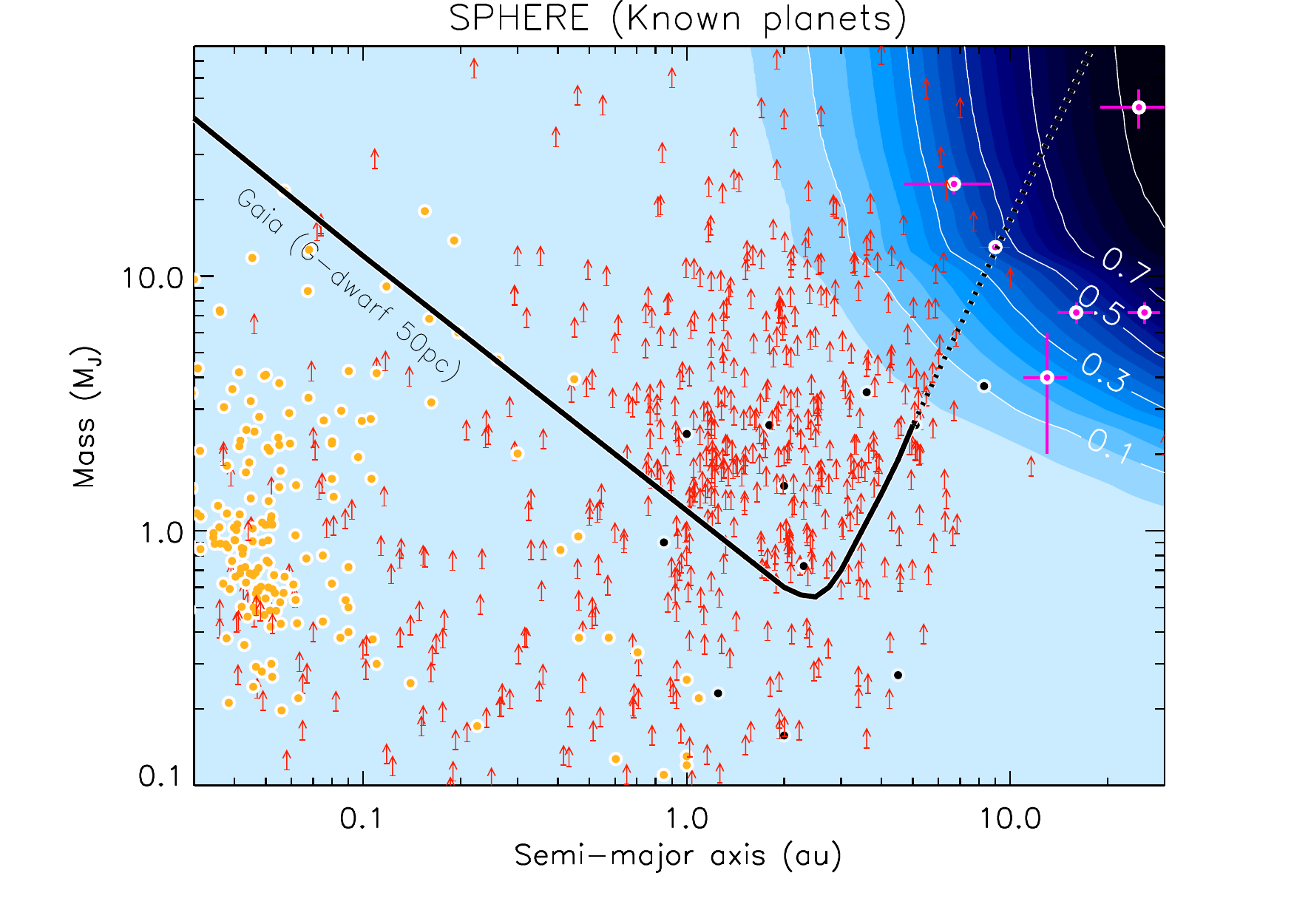}
\includegraphics[height=5.5cm]{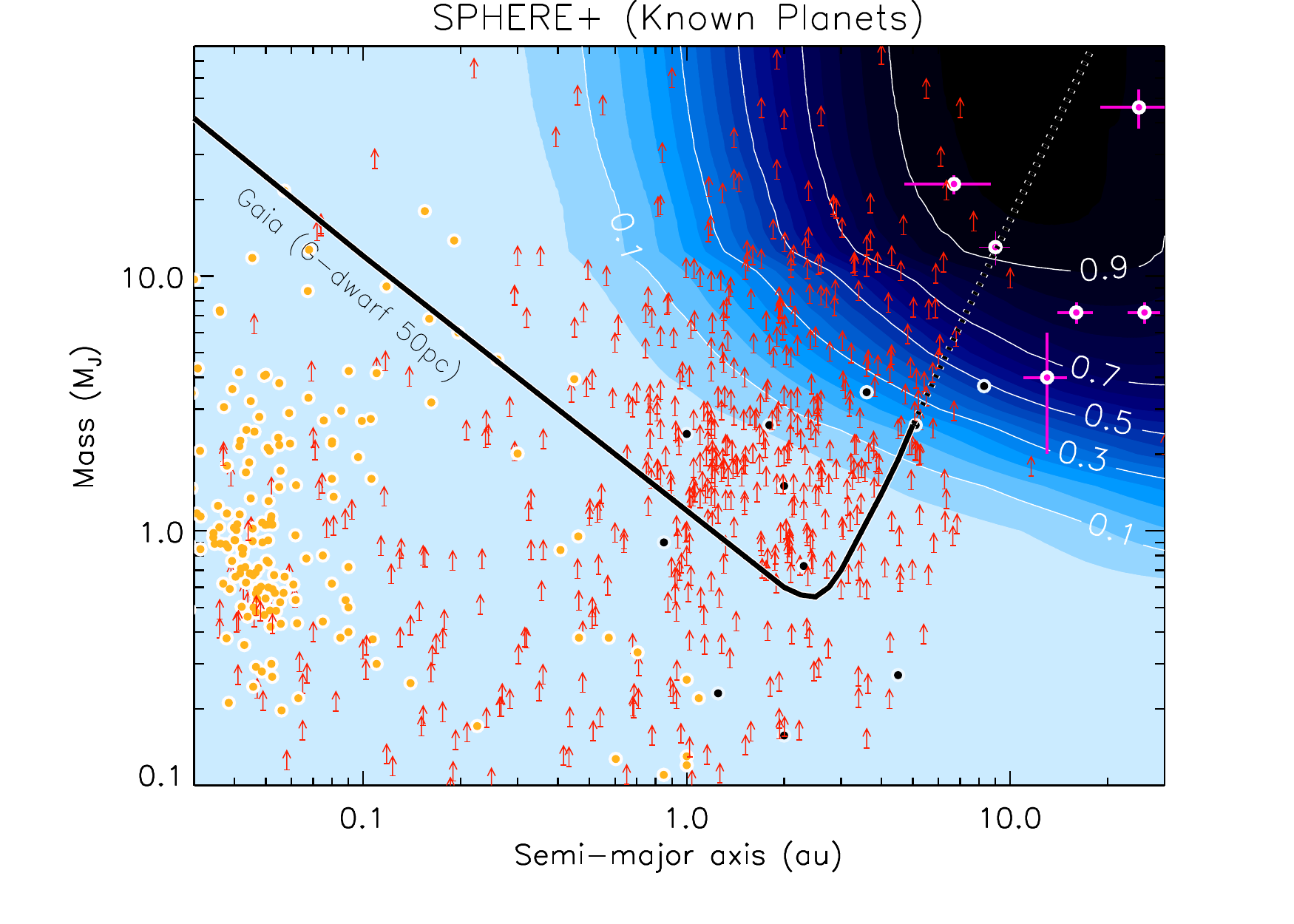}\\
\caption{SPHERE (\textit{Left}) and expected SPHERE+ (\textit{Right}) detection probabilities from the SHINE sample of young ($\sim 100$\,Myr), nearby ($\sim 50$\,pc) stars compared to the current population of exoplanets detected with all techniques: transit (\textit{yellow} dot), radial velocity (\textit{red} arrow), $\mu$-lensing (\textit{black} dot), and direct imaging (\textit{pink} dot). The \textit{Gaia} detection limits that will be available with the Data Release 4 in 2022 for a solar-type star at 50\,pc are overlaid. }
\label{fig:bridge}
\end{center}
\end{figure*}

For a sample of young, nearby stars (see Figure\,\ref{fig:shinestat}), SPHERE's current detection performance enables routine exploration of the population of giant planets located between 10 to 300 au and with masses $>$ 2\,$M_{\rm{Jup}}$. This parameter space has been extensively covered in the early statistical results from the SHINE campaign; however, the rate of new exoplanet discoveries remains low despite the large number of stars observed, conclusively demonstrating that the bulk of the massive giant planet population (observed via radial velocity, and located between 2\,au) does not extend significantly beyond 10\,au. The overlap between radial velocity (and soon also astrometry with \textit{Gaia}) and direct imaging remains currently marginal \citep{Lannier2017,Lagrange2018}. \textbf{A prime scientific goal of SPHERE+ will be to bridge the gap with indirect techniques by imaging young Jupiter down to the snowline at about 3\,au} (driving the TLR on the contrast performances in Table\,\ref{tlr1}). This will offer the unique possibility to obtain a complete view of the demographics of the giant planet population at all separations (see Figure\,\ref{fig:bridge}). 

While exoplanets are rare at separations $>$10\,au, radial velocity results suggest that significant populations of planets may be imageable at separations between 3-10\,au. 
As an illustration,  S\'egransan et al. (private com.) use the 22 years-long, volume limited, radial velocity CORALIE survey of nearby solar type stars to derive the companion mass distribution from the population of old giant planets to the lowest mass stars (see Figure\,\ref{fig:coralieMdist}, \textit{Left}). As expected, they find a decrease in the occurrence of giant planets as the mass of the planet increases up to  7\,M$_{\rm Jup}$, but also a significant increase in the distribution that peaks at 10\,M$_{\rm Jup}$ and slowly decays up to 30\,M$_{\rm Jup}$. The CORALIE survey indicates the existence of a distinct population of massive giant planets, not predicted by current core accretion scenarios \citep{Mordasini2017}, representing a minimum of 25 massive ($\ge2$\,M$_{\ Jup}$) planets located beyond 3\,au for a total of 1000 stars observed. \textbf{By reaching closer to the star and obtaining deeper contrasts (\texttt{tech.req.1}), SPHERE+ will be here unique to open a new parameter space between 3-10\,au to systematically target this population of massive giant planets at young ages, when they are in the early stags of formation and evolution, and to identify the peak and turnover after which the giant planet occurrence rate decreases with distance from the star.}

In addition, by the end of 2024 ($\pm0.5$\,years), the \textit{Gaia} astrometric survey will provide a complete census of all giant planets and brown dwarfs of the solar neighborhood orbiting stars of different masses and ages, and up to separations of 5\,au. As shown in Figure\,\ref{fig:bridge}, beyond 5\,au, the \textit{Gaia} performances are degraded but will still provide crucial information about the presence of massive giant planets and sub-stellar companions at several tens of aus. In the meantime, the \textit{Gaia} data release 4 (expected $\simeq$ 2022) will deliver 5.5\,years of fully calibrated astrometric time series that we will use to search for indications of the presence of giant planets at large separations. Building and exploiting SPHERE+, at the era of \textit{Gaia}, will significantly increase the number of discoveries of young massive Jupiters down to the snowline. \textbf{SPHERE+ will further characterize the physical properties (luminosity, atmosphere, orbit, interactions with the environment) of these young Jupiters located between 3-10\,au}.

Consequently, for this specific science case, SPHERE+ must achieve higher contrasts closer to the star in order to 
carry out dedicated programs for: 
\begin{itemize}
    \item Imaging and characterizing the young giant planets that will be discovered ($\leq5$\,au) or suspected due to astrometric trends ($\geq5$\,au) with \textit{Gaia} Data Release 4 (2022) and in the course of radial velocity campaigns targeting young, nearby stars with HARPS/3.6m, NIRPS/3.6m, ESPRESSO at VLT. \textit{Gaia} will observe 450 (and 1200), young ($\le500$\,Myr) stars at distances  closer than 50\,pc (and 150\,pc, respectively).
    
    \item Systematic surveying of the young, nearby stars observed during the SHINE campaign but at closer physical separations to explore the demographics of giant planets down to 3 au, in particular, with a deeper exploration between 3 to 20\,au, a parameter space that is only partially explored with SPHERE (and other techniques). SPHERE+ will serve here as a survey instrument providing further statistical constraints on the occurrence and distributions of giant planets in overlap with complementary techniques as radial velocity and astrometry. 
    These efforts will identify the most interesting young, planetary systems and exoplanets for further in-depth characterization with SPHERE+ itself, ERIS, Gravity/Gravity+, \textit{JWST} and the ELT.  As the ELT will not devote a significant number of nights for large-scale systematic surveys, it is vital now to reduce the sample of potential ELT targets and identify the most important systems for in-depth followup.  Carrying out this survey in service observing mode will be particularly important in order to take advantage of the best seeing and coherence time conditions and thus fully optimize the exploitation of SPHERE+.  

\end{itemize}

%---------
%---------

%%%%%%%%%%%%%%%%%%%%%%%%%%%%%%%%%%%%%%%%%%%%%%%%%%%%%%%%%%%%%%%%%%%%%%%%%%%%%
\subsubsection[Measuring the mass - luminosity of young Jupiter for the first time!]{Measuring the mass - luminosity of young Jupiter for the first time! (\texttt{sci.req.1})}
\label{sec:masslum}

Building on the synergy of SPHERE+ with other techniques like radial velocity and astrometry, a fundamental scientific goal beyond the discovery of young giant planets down to the snow line is the {\bf accurate determination of their physical properties}, which remains highly uncertain at young ages. Direct imaging yields photometry, luminosities and the spectral energy distributions of exoplanets, but does not directly measure exoplanet masses. 

%---------
\begin{figure*}[t]
\begin{center}
\includegraphics[height=5.5cm]{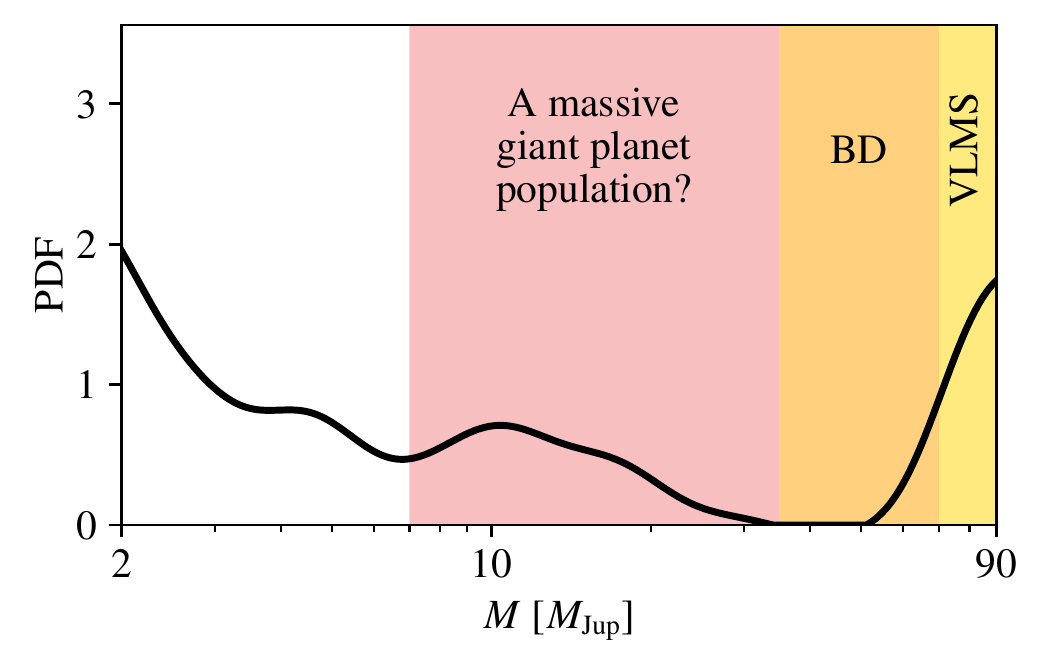}\hspace{0.5cm}
\includegraphics[height=5.5cm]{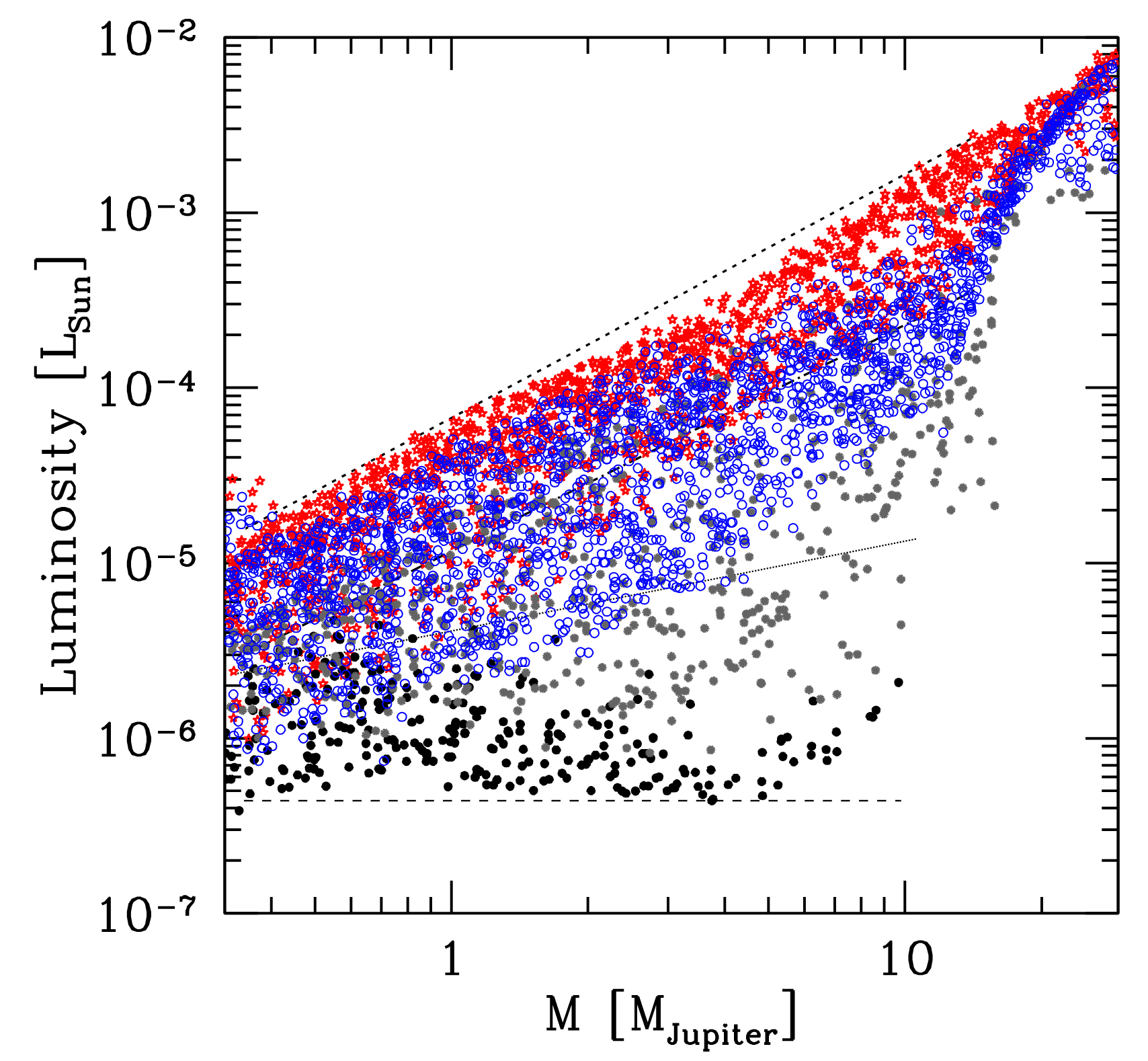}
\caption{\textit{Left,} Mass distribution of companions to solar type stars with semi-major axis smaller than 5\,au corrected from the $\sin i$ effect from the 22 years-long,  volume limited, radial velocity CORALIE survey of nearby solar-type stars. \textit{Right,} mass-luminosity relation  of  giant  planets  at  the  moment  when  the  protoplanetary  disk  disappears and predicted by current state-of-art models of planetary formation (here the BEX models, \citealt{Mordasini2017}). The blue empty circles are the so-called cold-nominal population while the red stars show the hot population. The black and gray dots show the cold-classical population.  
}
\label{fig:coralieMdist}
\end{center}
\end{figure*}
%---------

As directly imaged exoplanet masses have not been measured, we must rely on mass predictions from evolutionary models that are not calibrated at young ages \citep{marley2007}. In addition to the system age uncertainty, the predictions highly depend on the formation mechanisms and the gas-accretion phase that forms the exoplanetary atmosphere. Whether the accretion shock on the surface of a young accreting protoplanet is sub- or super-critical during the phase of gas runaway accretion will drive the initial entropy or internal energy, and hence strongly influence its initial physical properties (luminosity, effective temperature, surface gravity and radius) and their evolution with time \citep{marleau2014}. These different physical states are described by the so-called hot-start (sub-critical shock), cold-start (super-critical shock), and warm-start (intermediate case) models. These models predict luminosities that spread over several orders of magnitudes for young, massive giant planets. This is illustrated in Figure\,\ref{fig:coralieMdist} (\textit{Right}) showing the predictions of the BEX models from \citet{Mordasini2017}. Note the significant spread in post-formation luminosities of more than two orders of magnitude at a given mass. The higher the mass (up to 10\,M$_{\rm Jup}$), the larger the spread and the time duration for the models to converge. Nowadays, most masses reported in the literature for imaged exoplanets are the ones predicted by hot-start models and may under-estimate the planet masses as they fail to capture the accretion processes at play for the formation of giant planet atmospheres. \textbf{In concert with RV surveys and \textit{Gaia}, SPHERE+ will enable a joint determination of the dynamical mass, the luminosity and the spectral energy distribution of young Jupiters at given ages to calibrate for the first time their mass-luminosity relation}. This will offer the exclusive opportunity \textbf{to test evolutionary model predictions and study in more detail the formation histories of young Jupiters}, particularly the gas accretional heating and planetesimal accretion rate during the runaway phase of planet formation.

\begin{figure*}[t]
\begin{minipage}[c]{9cm}
\includegraphics[height=6.5cm]{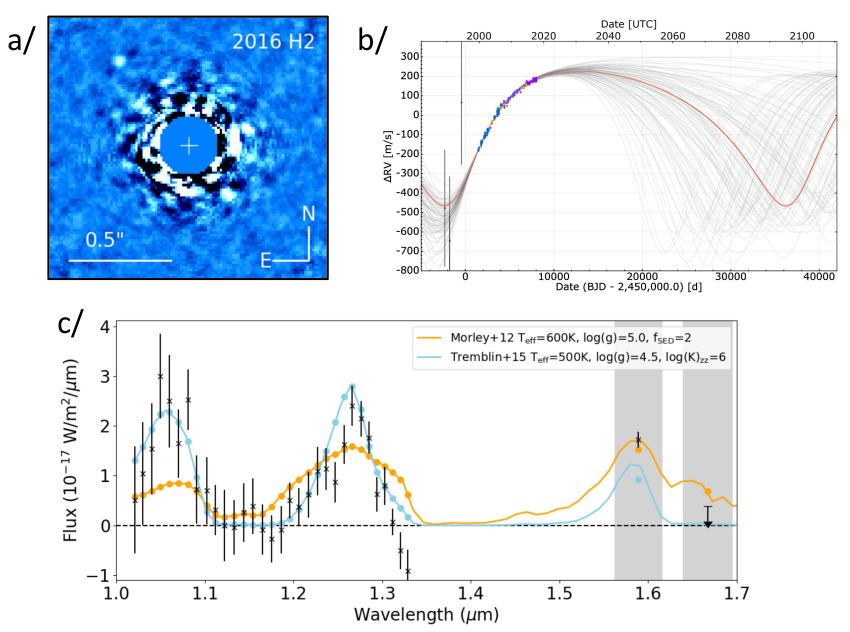}
\end{minipage}
\begin{minipage}[c]{8cm}
\caption{\textit{a/} SPHERE image of HD 4113\,A and C. \textit{b/} Radial velocity   measurements   of   HD 4113\,C   taken with CORALIE, HIRES and CORAVEL over a time baseline of 28\,years. \textit{c/} Comparison of the observed spectrum of HD 4113\,C with best-fit spectra from \cite{morley2012} and \cite{Tremblin2015} models. 
  }
\label{fig:ultracool}
\end{minipage}
\end{figure*}

The synergy between different detection techniques can be applied as well to the spectral and atmospheric characterization of ultracool ($\le600$\,K) brown dwarfs, which serve as benchmarks for both evolutionary models and atmospheric models at old ages. Figure\,\ref{fig:ultracool} shows an example of such a combined study for the T9-type brown dwarf companion HD\,4113\,C imaged around the star HD\,4113\,A, which itself hosts a 1.6\,M$_{\rm Jup}$ planet orbiting at 1.3\,au \citep{cheetham2018b}. By combining the observed astrometry from the imaging data with 28\,years of radial velocities, the analysis yields a dynamical mass of $66\pm4$\,M$_{\rm Jup}$ and moderate eccentricity of $0.44\pm0.08$, contradicting the isochronal mass estimate for this object and other objects with similar temperatures, even at the comparatively old age for this system of $5.0_{-1.7}^{+1.3}$\,Gyr. This result gives a first glimpse of what will be possible with SPHERE+ in combination with radial velocity and astrometric (with \textit{Gaia}) measurements; SPHERE+ in combination with complementary techniques will be able to measure the full mass-luminosity relation for young Jupiters and by extension for ultracool brown dwarfs.

%---------
%---------

%%%%%%%%%%%%%%%%%%%%%%%%%%%%%%%%%%%%%%%%%%%%%%%%%%%%%%%%%%%%%%%%%%%%%%%%%%%%%
\subsubsection[Exploring the diversity of young planetary atmospheres]{Exploring the diversity of young planetary atmospheres (\texttt{sci.req.3})}
\label{sec:spectro}

More than a decade ago, the first exoplanetary atmosphere studies focused on characterization of hot and strongly irradiated Jupiters like HD 209458 b \citep{Charbonneau2002} using transit observations.  Since then, transit spectroscopy observations have now been reported for roughly 100 exoplanets to date, including hot Jupiters, hot Neptunes, and super Earths (see \citealt{madhusudhan2019} for a recent review of the field). Currently, three main observing techniques are used to study the atmosphere of exoplanets: (i) low/medium resolution transmission (during transit) and emission (on secondary eclipse) spectroscopy (e.g., \citealt{deming2013}); (ii) high-resolution Doppler spectroscopy (\citealt{birkby2018} and references therein); (iii) high-contrast low-resolution spectroscopy (e.g., \citealt{Zurlo2016}). The last technique has been used in direct imaging to acquire the first spectra of non-strongly irradiated and self-luminous Jovian planets in wide orbits.  Characterizing young, self-luminous planets which are either in the process of forming or newly formed enables the detailed physical and chemical characterization of their atmospheres while also linking back to the formation mechanisms which might have formed such planets \citep{Chauvin2004,Janson2011, Konopacky2013, Bryan2018}. 

Low spectral ($R_{\lambda}=30-50$) resolution observations with the SPHERE Integral Field Spectrograph constrains the basic atmospheric properties of the planets, like the effective temperature, but fits of atmospheric models to low resolution spectra yield significant degeneracies in the determination of additional parameters like the surface gravity, the metallicity, the composition, or the presence of clouds \citep[e.g.][]{Bonnefoy2016}. The latest generation of models incorporate the best of our knowledge of cool planetary atmospheres to produce realistic synthetic emission spectra of planets. These models define the boundary conditions of the evolutionary models and are key to predict absolute magnitudes in given filter passbands (that are more accessible than bolometric luminosities). Most atmospheric models account for the formation of cloud particles of different composition (e.g., silicates, sulfites) and sizes. Clouds are a critical but complex component of the models, modifying the atmospheric structure (pressure-temperature profile), opacity, and composition (depletion of chemical elements). Each family of atmospheric models utilizes its own prescription for cloud physics and additional ingredients such as non-equilibrium chemistry and thermochemical instability. \textbf{With SPHERE+, accessing higher ($R_{\lambda}\ge5000$, \texttt{tech.req.3}) spectral resolution will resolve current degeneracies and fully explore the physics of the latest generation of planetary atmospheric models.}

\begin{figure}[t]
\begin{center}
\includegraphics[height=5.cm]{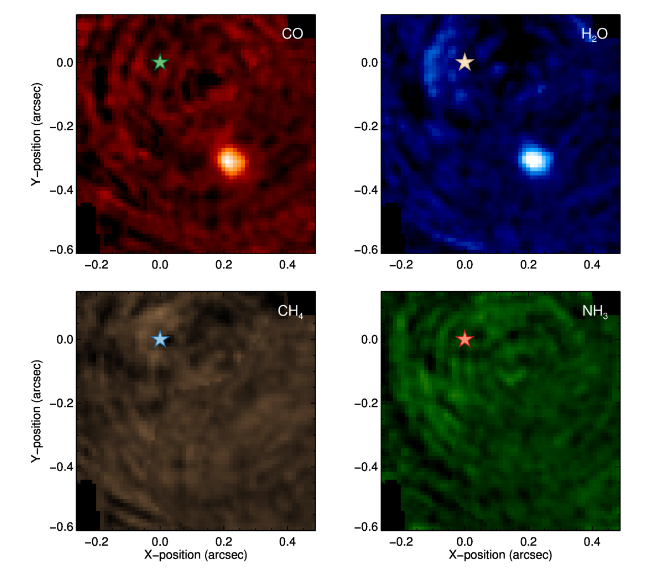}\hspace{0.5cm} 
\includegraphics[height=5cm]{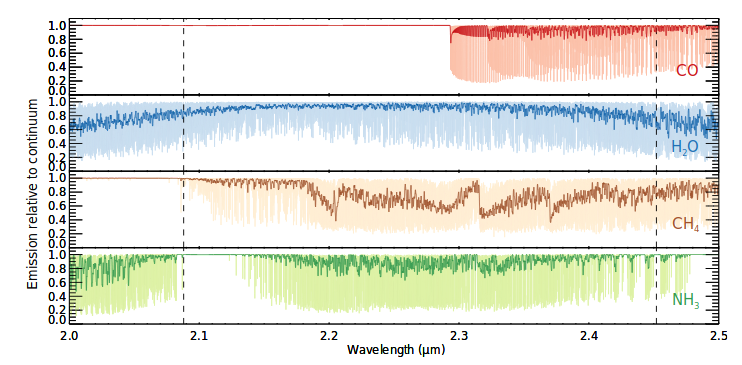} 
\caption{\textit{Left,}  VLT/SINFONI Molecule maps of CO, H$_2$O, CH$_4$ and NH$_3$ at v$_{\rm sys}=0$ km/s of $\beta$ Pic b. A cross-correlation enhancement caused by the planet is detected at S/N ratios of 14.5 and 17.0 in the maps of CO and H$_2$O respectively, but not in CH$_4$ and NH$_3$. \textit{Right,} Model templates of CO, H$_2$O, CH$_4$ and NH$_3$ at high spectral resolution of $R_\lambda=1\,000\,000$ (light colour) and convolved to a spectral resolution of $R_\lambda=5000$ (dark colour). The vertical dashed lines indicate the wavelength range of the data \citep[see][]{Hoeijmakers2018}.}
\label{fig:molmap}
\end{center}
\end{figure}

At higher spectral resolutions, the intrinsic spectral features of the planets can also be used to improve the detection sensitivity. This is partly used in the spectral differential imaging technique in combination with angular differential imaging (ADI), but in fact this technique relies more in the similarity of the speckles at different wavelengths than on the spectral features of the planet, although some detection limits estimations do use spectral templates to estimate the sensitivity to different spectral types \citep[][]{Marois2014, Ruffio2017}. To truly benefit from the very different spectral features expected for stars vs. planets, resolutions higher than at least $1000$s (ideally $10\,000$s) are necessary. Several characteristic examples have demonstrated the power of medium and high-resolution spectroscopy for the detection 
\citep[$\beta$ Pic b with VLT/SINFONI shown in Figure\,\ref{fig:molmap}, PDS 70 b and c,][]{Hoeijmakers2018, Haffert2019} or characterization 
\citep[$\beta$ Pic b with VLT/CRIRES, HR8799c with Keck/OSIRIS,][]{Snellen2014, Konopacky2013} of known companions. In such cases, the correlation of the extremely faint planetary signal with pre-computed spectral templates significantly boosts the detection sensitivity. Even in extremely noisy data strongly affected by stellar photon noise or detector noise, the planetary signal can be isolated and retrieved via a cross-correlation function. \textbf{At such a high spectral resolution, SPHERE+ will directly measure the radial and rotation velocities young Jupiters, leading to a better determination of the orbital properties, and even their rotational spin (for known inclination).  This will open up the unique opportunity to explore the spin velocity evolution of exoplanets with mass and age.} \vspace{0.1cm}

Therefore, SPHERE+'s enhanced characterization capabilities covering $J$, $H$ and $K$-bands (with TLRs summarized in Table\,\ref{tlr1}) will enable novel programs: 
\begin{itemize} 
\item to study for instance the accretion processes at play during the formation of giant planets, to measure the composition and abundances of specific elements tracing the origin of formation like carbon and oxygen, and to characterize the cloud properties (composition and size distribution) and both their horizontal and vertical spatial distribution, 
\item to resolve individual molecular lines (CO, CH$_4$, H$_2$O, NH$_3$) and measure the radial and rotational velocities of exoplanets, by accessing resolutions of at least $R_{\lambda}=5000$ and up to $100\,000$. 
\item finally, to explore the dynamics of the atmosphere during the object's rotation period and to actually map the 3D structures and composition of the exoplanetary atmosphere via temporal variability monitoring with SPHERE+'s increased stability.  Such variability studies provide key constraints for atmospheric models of these atmospheres, and will pave the way for future transformative studies with ELT instruments (like HARMONI and METIS), including possible Doppler Imaging of young giant exoplanets.
\end{itemize}

%%%%%%%%%%%%%%%%%%%%%%%%%%%%%%%%%%%%%%%%%%%%%%%%%%%%%%%%%%%%%%%%%%%%%%%%%%%%%
\subsubsection[Imaging young Jupiters in formation]{Imaging young Jupiters in formation (\texttt{sci.req.1, sci.req.2, sci.req.3})} 

ALMA high angular resolution observations of protoplanetary disks have revolutionized our view of disk evolution and shown that small-scale structures such as concentric rings are ubiquitous \citep[e.g.,][]{andrews2018}. Although the connection of these structures to planet-disk interactions is not definitely proven \citep{Flock2017, Bae2018}, it strongly suggests that planet formation might occur very early in the history of a young stellar system.

Current direct imaging surveys reach detection limits of a few Jupiter masses, but are often limited by bright and complex disk features.  Numerous claims of companion candidates in disks that show asymmetric features are indeed still debated or even rejected (e.g., HD\,100546, HD\,169142, LkCa\,15).  The recent discovery and confirmation of two young, dusty 7\,M$_{\rm{Jup}}$ planets in the young  transition disk surrounding PDS\,70 (see Figure\,\ref{fig:disk_rp_histogram}, \textit{Right}) opens the possibility for a more systematic search for and characterization of young proto-planets with SPHERE+, in both young star-forming regions as well as in the highly-structured planet forming disks observed with ALMA \citep{Keppler2018,Keppler2019}. Enabling such systematic searches for young proto-planets with SPHERE+ drives two key top level requirements (TLRs) for the instrument, reported in Table\,\ref{tlr1}. The second-stage AO proposed for SPHERE+ will enable a much higher image quality than is currently achieved due to its higher operational frequency ({\texttt{tech.req.1}}). This will have a major impact on differential imaging techniques, in particular reference differential imaging and molecular mapping, which will disentangle young planets and disk structures. Increased wavefront sensing capabilities will enable observations of faint and red stars ({\texttt{tech.req.2}). With such upgrades, \textbf{SPHERE+ will be unique in its ability to systematically survey the environment of young stars in star-forming regions (out of reach of current xAO planet imagers) and to further discover young Jupiters in formation and PDS\,70 system analogs} (and potentially even younger examples) by significantly increasing the sample size of observable targets (see Figure\,\ref{fig:disk_rp_histogram}, \textit{Left} and also Figure\,\ref{fig:saxo-sensitivity}). 

%---------
\begin{figure}[t]
\begin{center}
\includegraphics[height=6cm]{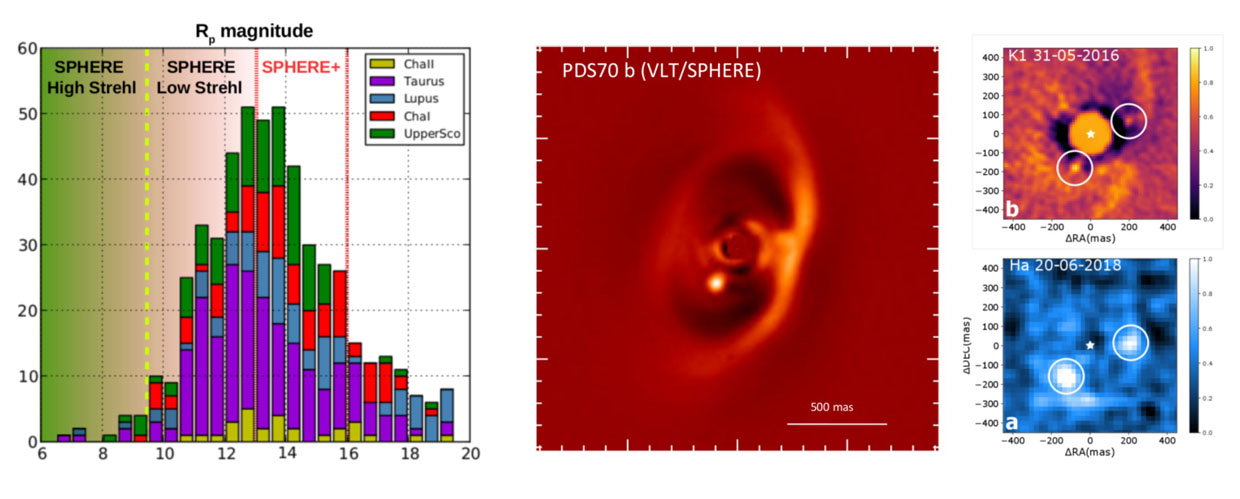}%\hspace{0.5cm}
\caption{\textit{Left,} Histogram of young members of nearby star-forming regions, prime targets of ALMA observing campaigns including the DSHARP survey. SPHERE's high-Strehl correction cutoff is located at $R=9.5$. \textit{Right,} SPHERE image of the PDS70 exoplanetary system together with a zoom-in on the planets b and c seen by SPHERE at $K_1$-band and by MUSE in $H_\alpha$ \citep{Muller2018,Haffert2019}.}
\label{fig:disk_rp_histogram}
\end{center}
\end{figure}
%---------

Accretion markers (mainly H lines here) play a crucial role in the detection of embedded protoplanets. Young accreting planets and brown dwarfs can be observed in very narrow bands in lines tracing accretion such as H$_{\alpha}$ (but also Pa$_\beta$, Br$_\gamma$ in near-infrared). The huge potential of H$_{\alpha}$ investigation has been demonstrated by recent detections of actively accreting companions with \textit{HST} \citep[GQ Lup b and DH Tau b,][]{Zhou2014}, Magellan/MagAO \citep[HD 142527 B, PDS 70b,][]{Close2014, Wagner2018}, and more recently VLT/MUSE \citep[PDS 70b, and c,][]{Haffert2019}. For accretion markers, ZIMPOL has been capable of achieving unprecedented angular resolution ($\sim$20\,mas) in visible. However, the observed contrasts in H$_{\alpha}$ for all of these detections were of the order of $10^{-3}$. A high spectral resolution spectroscopic mode in SPHERE+ for H$_{\alpha}$ and/or Pa$_\beta$ will yield not only further order-of-magnitude gains in contrast, but also provide direct constraints on the mass accretion rate of the companions \citep{Aoyama2019}, in parallel to similar studies for low-mass young stars \citep[e.g.][]{Alcala2017}, and even on the mass of the planet (from the terminal velocity of the line profile). The higher the spectral resolution, the deeper is the contrast reached using accretion markers, as long as we are not limited by thermal background or read out noise from the detector. This implies that extremely deep searches for accreting objects are in principle possible provided that an IFU with a resolution of at least $R_\lambda\sim 10\,000$ is available (\texttt{tech.req.3}). This is precisely the science case for the near-infrared and visible integral field spectrographs proposed for the SPHERE+ upgrade, which will achieve $R_\lambda\sim 10\,000$ for a small bandwidth centered around the H$_{\alpha}$ and Pa$_\beta$-line. Therefore, \textbf{SPHERE+ will directly explore and characterize the physics of accretion processes and shocks during the planetary formation phase of young Jupiters.}

\vspace{-0.4cm}
\begin{centering}
\begin{table}[t]
    \centering
    \caption{Summary of the TLRs for the science case 3.1}\vspace{0.1cm}
    \begin{tabular}{|l|l|}
    \hline
    \hline
    xAO      & 90\% Strehl at $H$-band on bright ($R<9.5$\,mag) and red ($J<11$\,mag) targets \\
	         & 3kHz correction frequency for high PSF stability\\
    \hline
    Contrast (5$\sigma$) & $10^{-5}$ at 100\,mas (Goal:  $10^{-5}$ at 60\,mas) \hspace{1cm}\\
        \hline
    Spectral range & $R$, $J$, $H$ and $K$-bands \\
    \hline
    Spectral resolution & 50, $5\,000-10\,000$, $50\,000-100\,000$\\
    \hline
    Other & Pupil-tracking, NCPA correction, coronagraphy, Low-Res IFS\\
          & Medium-Res IFS in NIR (accretion lines, molecular mapping)\\
          & High-Res point-source in NIR (abundances, radial/rotational velocities) \\ 
          & Medium-Res for high contrast H$_\alpha$ imaging (accretion)\\
    \hline
    \end{tabular}
    \label{tlr1}
\end{table}
\end{centering}

%%%%%%%%%%%%%%%%%%%%%%%%%%%%%%%%%%%%%%%%%%%%%%%%%%%%%%%%%%%%%%%%%%%%%%%%%%%%%
\subsection{Planet forming disks and architectures}% ABo, CGi, FMe, MBe, CDo
\label{sec:disks}

\subsubsection{Imaging the bulk of the disk population}

The formation scenarios which can produce the diverse collection of planetary systems observed to date depend on the initial conditions in their parent disks (e.g. \citealt{Mordasini2012c, Mordasini2012b, Mordasini2012a}). Therefore to understand planet formation we must study the birth environments of nascent planets. It is particularly important to understand the timescales and pathways of disk evolution within the first 10 Myr, i.e. the time over which planets are thought to form.

\begin{figure}[b]
\begin{minipage}[c]{9cm}
\includegraphics[height=6.5cm]{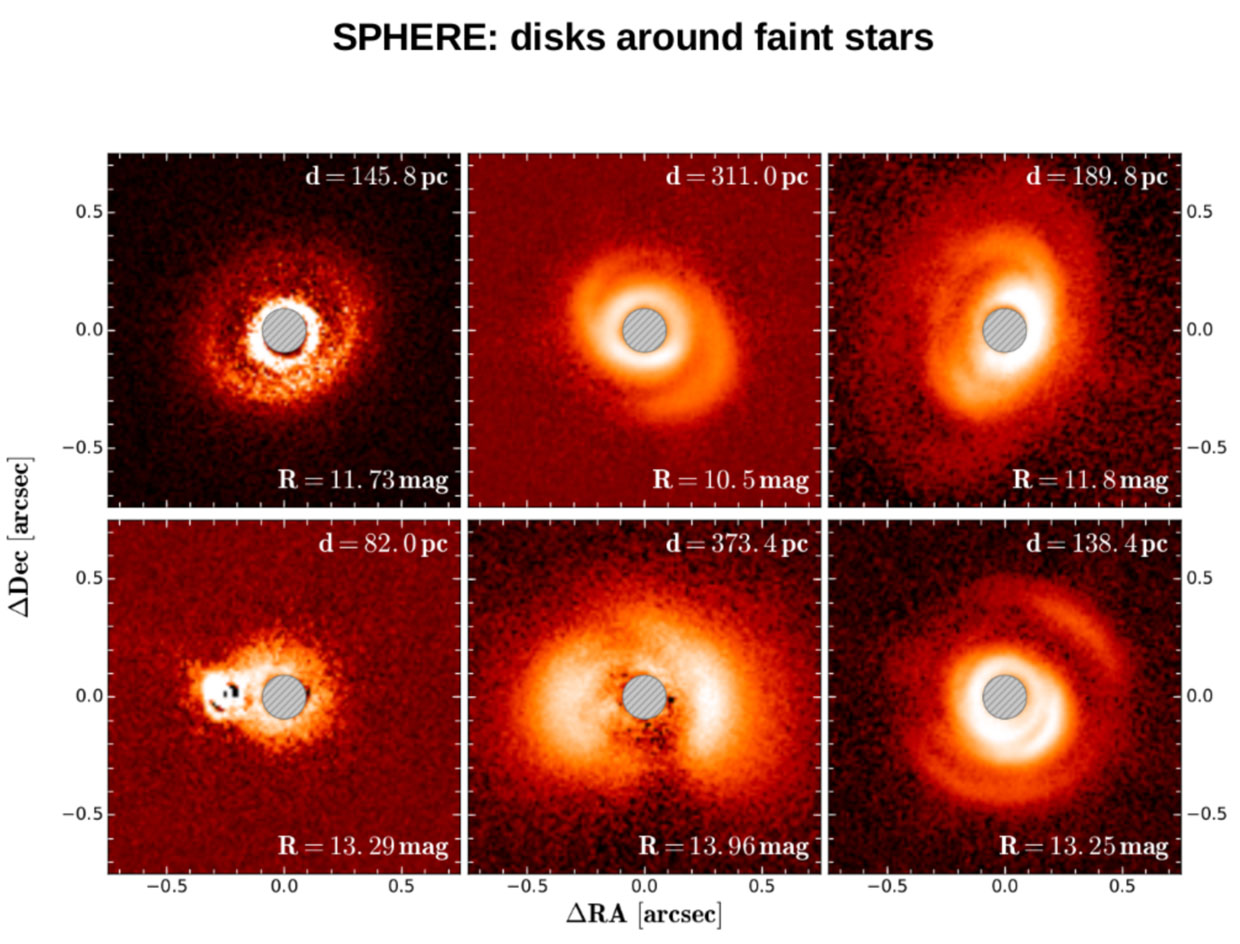}
\end{minipage}
\begin{minipage}[c]{8cm}
\caption{low mass (and thus faint) T Tauri stars observed with SPHERE in exceptional conditions. These observations illustrate the wealth of information gained from (common) T Tauri stars (SPHERE consortium, in prep.). SPHERE+ will be able to observe such systems routinely and extend the parameter space to even lower stellar and dust masses.}
\label{fig:faint}
\end{minipage}
\end{figure}

Due to the limitations on target brightness imposed by the optical wavefront sensor of SPHERE's AO system, disk observations were primarily limited to Herbig AeBe stars and the brightest T Tauri stars, leading to the identification of the most massive disks around the most massive stars.  But how do these massive disks compare to disks around lower mass stars, the most common stars in the galaxy? In Figure~\ref{fig:disk_rp_histogram} (\textit{Left}), we show the optical $R$-band magnitudes of members of nearby star forming regions. Currently, SPHERE can only observe a small fraction of these systems -- most are too faint for the current AO architecture. To address this science case in an exhaustive way simply requires increasing the statistics for a better sampling of the parameter space in system age, stellar mass, disk size and mass. It is desirable to extend these parameters  towards younger ($<$\,1\,Myr) and older ($>$\,30\,Myr) ages, low mass stars (M type stars), as well as disks with smaller angular sizes ($<0.1^{\prime\prime}$) and fainter disks (contrast $> 10^{-5}/10^{-6}$), in order to cover the full evolutionary sequence. \textbf{SPHERE+'s xAO upgrade will enable the regular observation of faint, red targets with $R$-band magnitudes down to $\sim$16\,mag. This includes then the vast majority of all nearby young systems.} In Figure~\ref{fig:faint}, we show a few of the faintest stars that have been observed with the current version of SPHERE in excellent observing conditions, demonstrating clearly the invaluable information we will collect on the circumstellar environment of low-mass stars.  These observations, difficult and rare with SPHERE, would become routine with SPHERE+, \textbf{thus providing access to the bulk of the disk population.}

Disk science with SPHERE benefits both from a strong synergy with ALMA at sub-mm wavelengths and also in the near future with \textit{JWST} in the mid-infrared (despite a degraded spatial resolution).  SPHERE measures pure scattered light from the disk surface, while ALMA is sensitive to thermal emission from the mid-plane, and \textit{JWST} can detect a mix of both from intermediate heights in the disk. Multi-wavelength observations are thus vital to better characterize the spatial distribution of the dust, with each spectral range sensitive to different typical grain sizes and locations. \textbf{Together with ALMA and \textit{JWST}, SPHERE+ will crucially observe the same systems at different wavelengths and enable large surveys of the most common types of circumstellar disks.}

%%%%%%%%%%%%%%%%%%%%%%%%%%%%%%%%%%%%%%%%%%%%%%%%%%%%%%%%%%%%%%%%%%%%%%%%%%%%%
\subsubsection{How do planets interact with the disk?}

%---------
\begin{figure}[t]
\begin{center}
\includegraphics[height=5.5cm]{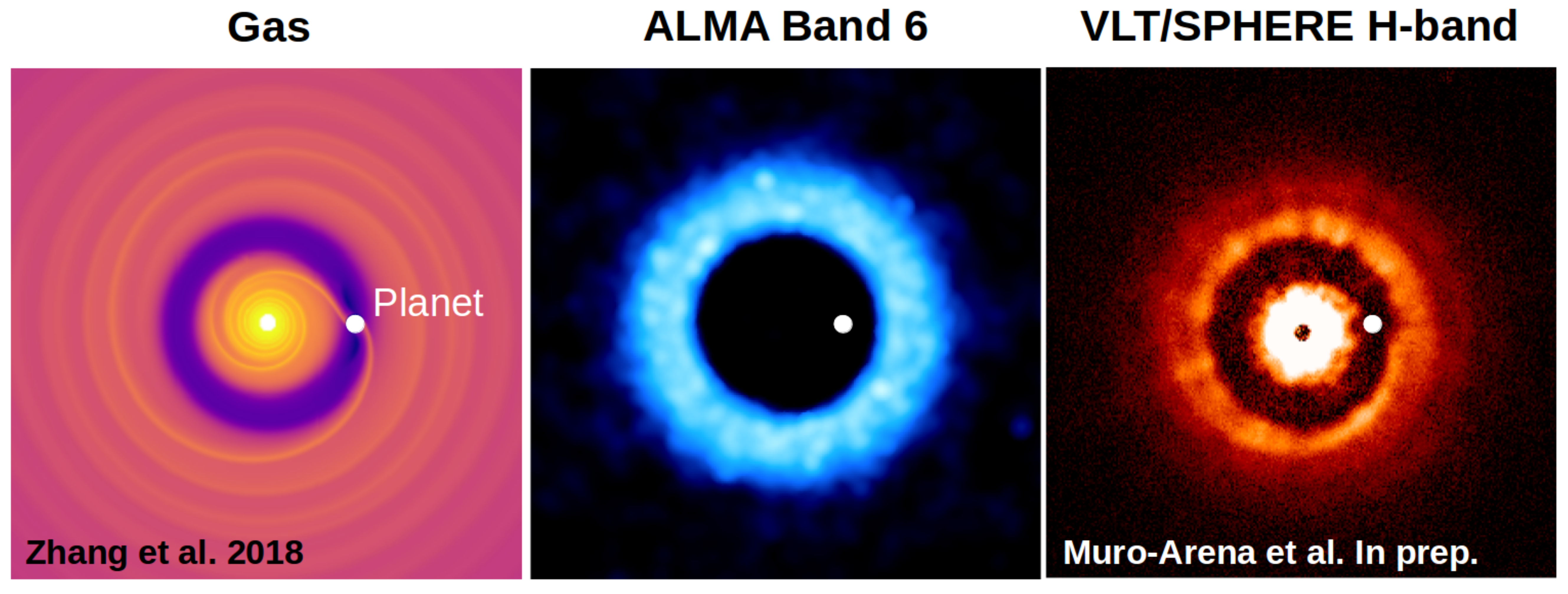}
\caption{Hydrodynamic model of a 3\,M$_{\rm Jup}$ planet in a gas rich disk by \cite{zhang2018} (left panel). We show simulated ALMA and SPHERE observations in the middle and right panels (Muro-Arena et al., in prep.). The planet position is indicated with a white dot. While ALMA only shows a ring with a cavity in dust continuum emission, SPHERE traces the small dust particles that follow the gas and reveal the asymmetric structure introduced in the gas by the planet.}
\label{fig:planet-disk-interaction}
\end{center}
\end{figure}
%---------

\textbf{To connect the known population of old planets with the young gas-rich disks in which they formed it is necessary to understand how planets interact with their parent disks.} In particular planets are susceptible to migration, as evidenced by the many planets found in radial velocity or transit at short orbital periods. These observations, however, are of old systems, where migration is already complete. In the course of their migration, planets can open gaps which will modify migration.  Imaging is the only way to constrain the locations and the time frame of such gaps. A systematic mapping of gaps with respect to system age within the first $10$\,Myr provides a method to witness migration in planetary systems since gap locations will follow the pressure bump introduced by the migrating planet. Here again, the comparison of SPHERE and ALMA images is crucial to understand how planetary evolution constrains the dust particle size distribution. Grain segregation in disk gaps and cavities can tell us what is causing the observed morphological features. 

The small ($\mu$m-sized) dust particles traced in scattered light are furthermore well coupled to the gas in the disk and can thus be used to trace pressure wakes induced by planets in the disk gas. In Figure~\ref{fig:planet-disk-interaction}, we show ALMA and SPHERE synthetic images derived from hydrodynamic simulations. While ALMA shows a smooth ring in continuum emission, SPHERE traces the asymmetric signature (in particular spiral waves) visible in the gas.  \textbf{The higher sensitivity wavefront sensor proposed for SPHERE+ is necessary to study these spatial substructures in disks around low mass stars. Additionally, higher contrast close to the star is required to study the area close to the ice-lines (corresponding to separations of a few au from the star).} For the typical distance to the closest star forming regions ($\sim 150$~pc), the projected separation is $\sim$100 mas or even less, separations at which ALMA is already revealing a great variety of structures in protoplanetary disks, all of which will be observable with SPHERE+ \citep[e.g.][see Figure~\ref{fig:alma_dsharp_kinematics}, \textit{Left}]{andrews2018}. 

Finally, ALMA is slowly starting to provide deep disk images tracing the CO gas content at high spectral resolution. Deviations from the expected Keplerian rotation patterns have led to the discovery of a few planets, interestingly all located in observed gaps in the continuum (\citealt{pinte2018, teague2019, pinte2019}, see Figure~\ref{fig:alma_dsharp_kinematics}, \textit{Right}). In addition to future PDS70b-analog discoveries in the course of SPHERE+'s systematic survey of star-forming regions, as ALMA keeps observing more systems, the number of kinematical signatures (and thus planets) will rise.  Thus, the synergy with ALMA to drive a dedicated SPHERE+ campaign for planet imaging is obvious: ALMA shows where to look for the planet, SPHERE+ will image and characterize the corresponding point source and its interaction with the surrounding disk.

%---------
\begin{figure}[t]
\begin{center}
\includegraphics[height=6.5cm]{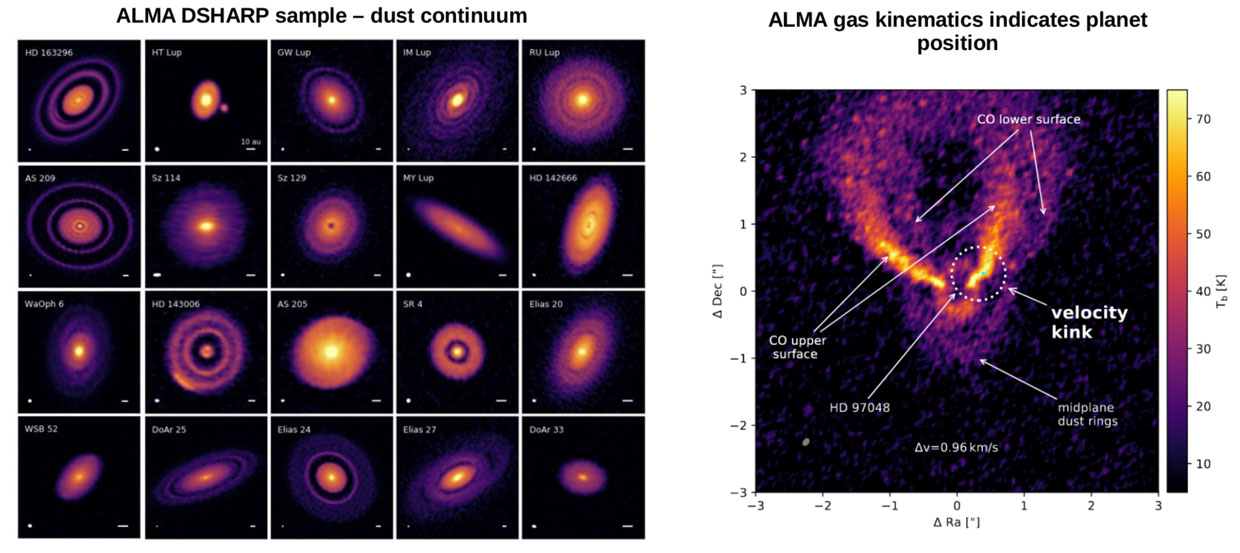}
\caption{\textit{Left,} dust continuum images of 20 young circumstellar disks taken as part of the DSHARP survey with ALMA at the highest spatial resolution available. Sub-structures are revealed in all disks.
\textit{Right,} velocity channel of the $^{12}$CO emission line in the HD\,97048 system that illustrates gas radial velocity. A localized kink in the iso-velocity profile is detected that indicates the presence of a giant perturber (\citealt{pinte2019}).}
\label{fig:alma_dsharp_kinematics}
\end{center}
\end{figure}
%---------

%%%%%%%%%%%%%%%%%%%%%%%%%%%%%%%%%%%%%%%%%%%%%%%%%%%%%%%%%%%%%%%%%%%%%%%%%%%%%
\subsubsection{How does the dust evolve?}
\label{sec:debris}

Due to gas drag, dust grains settle to the mid-plane and grow. This dust component is then incorporated into planet cores and planet atmospheres. Later on in a disk's evolution, the dust observed in debris disks results from collisions between planetesimals and hence can have different properties. \textbf{Investigating the variation of dust shape and size distribution as a function of stellar mass, and system age is key for understanding planet formation and the planetary environment.} SPHERE can characterize dust properties either from polarimetry or from photometry/spectroscopy or both. Resolved photometry or spectroscopy (delivering information at several phase angles) has been performed in a very few cases \citep[][]{Milli2017, Milli2019, Bhowmik2019} and already provides crucial details about dust grain size distribution either by measuring the scattering phase function (Figure \ref{fig:hr4796spf}) or the spectrum. The canonical picture of debris disks presumably populated with grains larger than the blow-out size is starting to evolve.  % added JMi
The combination of phase function, polarization fraction and spectral reflectance can break the degeneracies to reveal the dust shape, size and composition, providing vital information on the conditions of dust growth in the very first stages of planet formation. Particle shapes appear more complex than compact spheres, with phase functions suggestive of large and porous aggregates, consistent with in-situ measurements of solar system comets.

In addition, several ALMA observations provide evidence of gas (possibly 2$^{\rm nd}$ generation) in debris disks, which modifies the dynamical behavior of grains \citep{Kral2019}. Finally, the inner regions where warm dust is co-located with planets (even terrestrial) have been little explored in both planet-forming and debris disks. 
This region is of interest in order to understand how and when dust dissipates from the inner system and to connect interferometric observations of exozodii with direct imaging. 
Given the intensity variations of scattered light vary as the square of the stellocentric distance, we can expect these regions to be rather bright. With the proposed xAO upgrades 
\textbf{SPHERE+ will be uniquely capable of studying dust properties and mineralogy, and exploring the disk inner regions where telluric planets are expected to form for very nearby stars (possibly up to a distance of $\sim$50\,pc).}

\begin{figure}[t]
\begin{center}
\includegraphics[height=5.5cm]{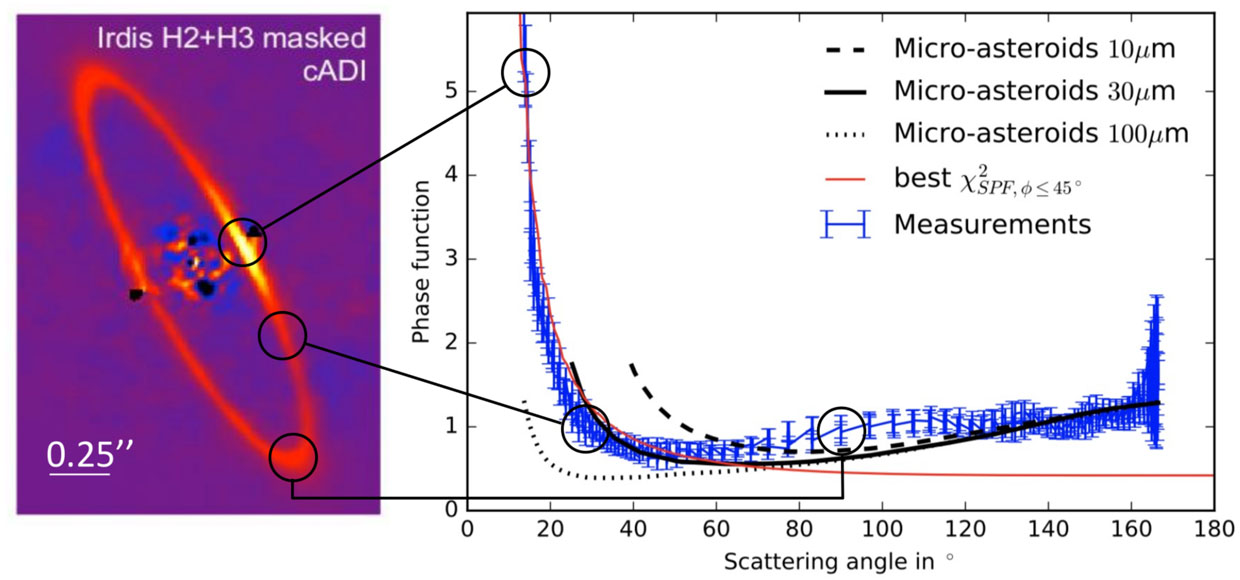}
\caption{Scattering phase function of the debris disk HR4796 \citep{Milli2017} compared to models of grains with various sizes showing how important are the small scattering angles (at the closest projected separations) to constrain the grain size distribution.}
\label{fig:hr4796spf}
\end{center}
\end{figure}

For this science case, TLRs are summarized in Table\,\ref{tlr2}. Connecting TLRs to the science case: i/  higher contrasts at 2-3$\times$ smaller separations are necessary for a full exploration of the warm dust content located at $<$5\,au, ii/ higher and more stable contrasts with little or no impact on flux from the data reduction (so-called self-subtraction in angular differential imaging) is required for the the measurement of the disk surface brightness (total or polarized) at the smallest scattering angles. This is relevant even at low resolution to enable thorough characterization of grain properties at different stages of the evolutionary sequence. Improving the polarimetric accuracy is also crucial to better characterize grain particles in comparison with ZIMPOL observations.

\vspace{-0.4cm}
\begin{centering}
\begin{table}[!h]
    \centering
    \caption{Summary of the TLRs for the science case 3.2}
    \begin{tabular}{|l|l|}
    \hline
    \hline
	         xAO      & 90\% Strehl at $H$-band on bright ($R<9.5$\,mag) and red ($J<11$\,mag) targets \\
	         & 3kHz correction frequency for high PSF stability (RDI)\\
    \hline
    Contrast (5$\sigma$) & $10^{-5}$ at 100\,mas (Goal:  $10^{-5}$ at 60\,mas) \hspace{1cm}\\
    \hline
    Polarimetry & de-rotator re-coating to minimize polarimetric beam shifts \\
		& (accessing the inner $\sim$ 100 mas with DPI observations in multiple bands)\\
		& polarizing beamsplitter for IRDIS for 50\,\% efficiency increase in large surveys\\
    \hline
    Spectral range & $R$, $J$, $H$ and $K$-bands \\
    \hline
    Other & Pupil-tracking (ADI), NCPA correction, Coronagraphy\\
    \hline
    \end{tabular}
    \label{tlr2}
\end{table}
\end{centering}

%%%%%%%%%%%%%%%%%%%%%%%%%%%%%%%%%%%%%%%%%%%%%%%%%%%%%%%%%%%%%%%%%%%
\subsection{Birth and death of stars}%  

\subsubsection{Jets and outflows from young stellar objects}%  %[S. Antoniucci, E. Regliaco, C. Dougados] 

Studies of YSO jets from both classical T Tauri stars and the more massive Herbig AeBe stars suggest that jets originate from an extended region of the disk and are accelerated and collimated by the action of a large-scale magnetic field, anchored in the rotating star-disk system on scales of fractions up to tens of au \citep[e.g.][]{frank_2014}. On this basis, jets are expected to play a major role in extracting the angular momentum from the system and have a deep influence on the properties of the disk region in which planets form. 

In this context, \textbf{observing the very base of the jets is therefore a crucial aspect for both assessing their magneto-centrifugal formation mechanism and determining the region where they originate, hence, their feedback on the disk}. This investigation requires probing jet tracers (such as H$_\alpha$, [OI], and [FeII], Pa$_\beta$) at distances smaller than $5-10$\,au from the driving source, which correspond to angular separations smaller than $35-70$\,mas for close-by ($\sim$140~pc) YSOs. SPHERE has recently provided the first direct insights on the launching region of YSO jets, through observations both in the optical and near-infrared of typical jet/wind tracers \citep{antoniucci2016, garufi_2019, rigliaco_2019}. 
From an observational point of view, the main limitation in studying the jet base is the ability to perform an optimal subtraction of the star continuum as close as possible to the central source. This requires reaching contrasts up to 10$^{-3}-10^{-4}$ at angular separations $<100$\,mas. In this respect, \textbf{SPHERE+ can provide a major breakthrough with its improved performance at short angular separations from the source}. With a spectral resolution of $R_\lambda=5000-10\,000$ delivered by an IFS in both visible and near-infrared, we would be able \textbf{to image the jet in separate velocity channels in each line} (jet line profiles are broad, typically $200-400$\,km/s). This would provide valuable information concerning the kinematics of the outflow at its base.  Measuring the kinematics of the outflow at its base is an important test of the magneto-centrifugal model, according to which the flow has an onion-like kinematic structure, with progressively slower components launched from disk regions at larger distance from the star.  Access to high contrasts at spectral resolutions of $R_\lambda=5000-10\,000$ would thus be huge improvement over the previous capability of SPHERE.  Such a capability would be ideally suited for thes study of YSO jets, as exemplified by the work conducted on the DG Tau jet with SINFONI at VLT with spectral resolution of $R_\lambda=4000$ \citep{agra-amboage_2011,agra-amboage_2014}. Observations in multiple lines will \textbf{offer the possibility of deriving the physical conditions of the gas (ionisation, electron density, temperature) in the inner few \,au} of the jet from line ratios and their variation with velocity, using dedicated and well-tested spectral diagnostic techniques \citep{bacciotti_1999,giannini_2019}. Finally, SPHERE+ might \textbf{provide a measurement of the rotation of the jets}, which is a fundamental parameter for the computation of the angular momentum removal, provided that  a high enough spectral resolution is available ($R_\lambda\sim 10\,000$). The related TLRs for this science case are reported in Table\,\ref{tab:jets_tlr}. 

\subsubsection{Evolved stars}%  %  [E. Lagadec, P. Kervella] 

\textbf{Evolved stars}, and particularly Asymptotic Giant Branch (AGB) stars and Red Supergiants (RSG), \textbf{are the main contributor to the chemical enrichment of galaxies. Understanding when and how material is ejected from evolved stars is necessary to build realistic stellar evolution models} \citep{hoefner2018}. There are converging indications that mass loss from evolved stars is powered by a combination of pulsation, convection, and radiation pressure on dust formed at a few stellar radii. To assemble a coherent picture of the dissemination of star-processed material, we are missing high precision observational constraints on the circumstellar material distribution and of its chemical evolution as a function of radius in the close environment of the stars (within $\approx 100$ stellar radii). TiO molecules veil the photospheres of these giant stars. This could be one of the explanations for the spectacular dimming of the RSG Betelgeuse in late 2019/early 2020. ZIMPOL observations of R\,Doradus have revealed spectacular variations of its surface shape that could be due to a MOLsphere, an extended gaseous atmosphere, containing TiO \citep{Khouri2016}.  An IFS in the optical will enable us to map such MOLspheres, but also H$_{\alpha}$ emission due to shocks, leading to dust formation. Finally, the increased polarimetric accuracy of SPHERE+ will allow us to map the dust distribution that is scattering light from the central star \citep[see, e.g., ][]{kervella2015, Kervella2016}, and to determine its size and composition, which is key to quantitatively study radiation pressure on these grains. 

With mass loss rates up to 10$^{\rm{-4}}$\,M$_{\odot}$\,yr$\rm ^{-1}$, AGB stars shed most of their mass in a few 10$^4$ years, so that this phase is very short. It is estimated that in the Large Magellanic Cloud, the 7$\%$ highest mass-losing stars contribute to 66$\%$ of the dust production \citep{Srinivasan2016}. Studying these extreme evolved stars is the key to studying galactic chemical enrichment. These rare stars are deeply embedded, due to their thick dusty envelopes. With its near infrared wavefront sensing capability, \textbf{SPHERE+ will enable us to map their circumstellar environment from the dust formation radius outwards}. SPHERE+ will thus provide key clues on both \textbf{the mass-loss mechanisms and history}. Finally, \textbf{binarity} impacts the yields from the stars, e.g., via mass transfer that modifies their chemical composition and structure. Nearby companions can also trigger envelope ejections, and thus eject gas before it is enriched in newly formed elements.  Ultimately, as shown with SPHERE for the prototypical dusty Wolf-Rayet star WR 104 \citep{soulain2018}, dust formation is due to the interaction of the wind of the WR star with its O type companion, forming a spectacular spiral. Only the closest dusty WR stars can be observed by SPHERE, as these objects are embedded. SPHERE+ will enable us \textbf{to map the environment of a dozen WR stars and thus better understand the dust enrichment of our Galaxy}. 

%Table YSOjets 
\vspace{-0.4cm}
\begin{centering}
\begin{table}[h]
\centering
\caption{Summary of the TLRs for the science case 3.3}
\begin{tabular}{|l|l|}
\hline
\hline

	         xAO      & 90\% Strehl at $H$-band on bright ($R<9.5$\,mag) and red ($J<11$\,mag) targets \\
	         & 3kHz correction frequency for high PSF stability (RDI)\\
\hline
Contrast  \hspace{0.5cm}   &   $10^{-3}$-$10^{-4}$ below 100\,mas \\
& goal: achieve best possible star subtraction at 35 mas (5~au at 150~pc)\\
& to observe jet/outflow base in close-by YSOs and evolved stars \\
\hline
Spectral resolution   & $R_\lambda\sim 5\,000-10\,000$ to resolve various velocity channels \\
			&  $R_\lambda \sim10\,000$ to measure jet/outflow rotation \\
\hline
Wavelength range    & VIS: $0.62-0.67\,\mu$m to cover H$\alpha$ and [OI]$\lambda$6300 \\
                                & NIR: $1.2-1.8\,\mu$m,  to cover Pa$\beta$, [FeII]1.25 and [FeII]$1.64\,\mu$m\\

\hline
\end{tabular}
\label{tab:jets_tlr}
\end{table}
\end{centering}

%%%%%%%%%%%%%%%%%%%%%%%%%%%%%%%%%%%%%%%%%%%%%%%%%%%%%%%%%%%%%%%%%%%
\subsection{Solar system}

\textbf{The discovery and dynamical characterization of moons of asteroids provides a key example of a prime solar system science case for SPHERE+.} 
Multiple-asteroid systems (binaries, triples) represent a sizable fraction of the asteroid population \citep{margot2015}. Furthermore, they enable investigations of properties and processes that are otherwise difficult to probe by other means. In particular, Earth-based observations of multiple systems provide the most productive way of deriving precise masses and hence densities for a substantial number of objects \citep[see, e.g., ][]{carry2012,marchis2013}. Measured densities are particularly vital to constrain the internal structure of these objects, as this density depends on the location and timing of their formation in the early solar system. As the only other way to constrain asteroid masses with similar precision is by dedicated interplanetary missions (either a fly-by or a rendezvous, e.g., ESA \textit{Rosetta}, NASA \textit{Dawn}, \textit{OSIRIS-Rex}, JAXA \textit{Hayabusa 1 \& 2}), direct imaging is a very efficient approach for discovering new moons and for constraining their orbital parameters, and hence the total mass of the system (primary + secondary). In the case of a small secondary (which is always the case for large asteroids), the total mass is dominated by the primary implying that the mass of the primary can be well constrained (usually with a $\leq10\%$ uncertainty). Density can then be calculated by combining the measured mass with the 3D shape of the asteroid, reconstructed from disk-resolved images \citep[see, e.g., ][]{2019A&A...623A...6F}.

SPHERE has already produced stunning results regarding the internal structure of large main belt asteroids \citep{hanus2020,carry2019}, and the discovery of new moons around known asteroids \citep{yang2016,pajuelo2018, vernazza2019b}. With increased xAO sensitivity performances \textbf{SPHERE+ will enable systematic searches for new moons around the largest asteroids (typically the first 200 main belt asteroids).} We expect a large number of moons to be discovered including additional moons in known binary systems.  SPHERE+ will provide the means to also explore their orbits and formation mechanism(s), leading to improved characterization of the asteroids they orbit and allowing us to use them as tracers of the events that occurred in the early formation of the solar system.

The increased spectral capabilities of SPHERE+ in near-infrared will offer the unprecedented opportunity \textbf{to explore the dynamical processes at play in the atmospheres of giant planets and their satellites.} For instance,  Titan, the largest satellite of Saturn, is a fascinating evolved object with a surface shaped by active geologic and climatic processes. It also possesses a thick and dynamic atmosphere dominated by N$_2$ and CH$_4$ with the presence of photochemical hazes and icy clouds. In the optical domain a significant fraction of solar photons reflected by the surface can escape the atmosphere through very narrow spectral windows in the infrared. Between 1.2 and 1.85\,$\mu$m the medium resolution spectrograph will give access to two transmission windows (1.27 and 1.59\,$\mu$m) and their wings with a spectral resolution of $R_\lambda=5000$. Conjugated with a spatial resolution of the order of 260\,km, we foresee two main science goals: (i) \textbf{the monitoring of the atmospheric haze and tropospheric CH$_4$ and C$_2$H$_6$ clouds}, and (ii) \textbf{the spectral characterization of large geomorphological units of the surface.} In the case of Titan, such observations will be perfectly complementary with Cassini's VIMS data set and the expected \textit{JWST} observations. Planetary objects of the outer solar system (Iapetus, Triton,...)} will benefit from SPHERE+, especially those covered by tholins (heterogeneous organic compounds with various degrees of nitrogen incorporation) that will be mapped and characterized by narrow band imaging and polarimetry in visible and near-infrared. TLRs of both science cases are given in Table\,\ref{tab:tlr_solar}.

\vspace{-0.4cm}
\begin{centering}
\begin{table}[!h]
    \centering
    \caption{Summary of the TLRs for the Science Cases 3.4}
    \begin{tabular}{|l|l|}
    \hline
    xAO  \hspace{0.5cm}   & sensitivity to faint targets (up to $J\le14$) \\
                          & 90\% Strehl ratio at $H$-band on bright ($J<9.5$\,mag) targets \hspace{0.5cm} \\
                          & 3kHz correction frequency for high PSF stability (deconvolution) \\ 
\hline
    Spectral coverage & $V$, $R$, $J$, $H$ and $K$ bands \\
    \hline
    Spectral resolution & $R_\lambda \sim 5000$ in NIR,  NB and BB imaging in VIS\\
    \hline
    Other & IFS in NIR, Polarimetry \\
\hline
    \end{tabular}
    \label{tab:tlr_solar}
\end{table}
\end{centering}
%%%%%%%%%%%%%%%%%%%%%%%%%%%%%%%%%%%%%%%%%%%%%%%%%%%%%%%%%%%%%%%%%%%
\subsection{Active galactic nuclei}

\textbf{Due to their small sizes (1000\,au to 10\,pc), the innermost components of AGNs are still poorly understood.} 
According to the unified model proposed by \cite{1993ARA&A..31..473A}, if a central engine (CE) in the form of an accretion disk orbits a supermassive black hole then likely a dusty, molecular torus will exist circling the CE. This torus would be responsible for blocking the light emitted by the CE when observed edge-on.  The exact morphology and extension of the dusty torus, its link to the hundred pc-scale molecular ring, the mechanism for matter to lose angular momentum and feed the CE, as well as the path that matter follows, remain several important open questions.
Near-infrared is especially well suited for studying this environment, since it is sensitive to the central hot dust core and its warm surroundings through the obscuring material that blocks most of the visible light. However, because of AO sensitivity, there have been only very few successful attempts using adaptive optics in the near-infrared, with the most detailed results on NGC\,1068 where a pc scale was indeed reached (e.g. \citealt{rouan2004} using NaCo and \citealt{Gratadour2015} using SPHERE) and on the nearby AGN Circinus (see \citealt{mullersanchez2006} using SINFONI). 

With SPHERE, near-infrared polarization observations offers the advantage to not be spoiled by the dazzling central core. For NGC\,1068, SHERE has revealed an elongated structure of $\sim60$\,pc diameter in the same direction as the molecular torus and tracing presumably the outskirt of the torus through a double scattering process \citep{Gratadour2015,grosset2018}. Finally, spectroscopic diagnostics are key to disentangle the stellar radiation from the dust emission and the CE blue emission as well as to put constraints on the dust temperature, age of stellar population or hardness of ionizing radiation \citep{2019A&A...629A..98V}.  \textbf{Polarimetric and spectro-imaging at the very high angular resolution} provided by an AO system with better performances in sensitivity, as proposed with SPHERE+ (see TLRs summarized in table\,\ref{tab:tlr_agns}), \textbf{will be a stepping stone to provide a detailed and unprecedented view at a pc scale of a significant sample of nearby AGNs}, opening the door to comparative studies between various types of AGNs.  Considering the WISE catalog \citep{2015ApJS..221...12S}, which offers the most complete AGN catalog based on IR emission, we estimate that SPHERE+ will be able to observe 38 objects with $J\leq9$ and 77 objects with $J\leq10$ (compared to only 2 with SPHERE). \textbf{Given the dramatic increase of AGNs that will be observable, SPHERE+ will be a complete game changer} for the high angular resolution exploration of their innermost components including the morphology, extension, and physical properties of the dusty torus surrounding the CE.

\vspace{-0.2cm}
\begin{centering}
\begin{table}[h]
    \centering
    \caption{Summary of the TLRs for the Science Cases 3.5}
    \begin{tabular}{|l|l|}
    \hline
    xAO  \hspace{0.5cm}   & 90\% Strehl ratio at $H$-band on red ($J<11$\,mag) targets \hspace{0.5cm} \\
                              & 3kHz correction frequency for high PSF stability\\ 
\hline
    Spectral coverage & $J$, $H$ and $K$ bands \\
    \hline
    Spectral resolution & $R_\lambda \sim 5000$ in NIR, NB and BB imaging\\
    \hline
    Other & Coronagraphy, Polarimetry\\ 
          & Medium resolution IFS to cover H$_2$, Br$_\gamma$, Pa$\beta$, [FeII]1.25 and [FeII]$1.64\,\mu$m\\ 
          & (exploration of the molecular and ionized gaz) \\ 
\hline
    \end{tabular}
    \label{tab:tlr_agns}
\end{table}
\end{centering}

%%%%%%%%%%%%%%%%%%%%%%%%%%%%%%%%%%%%%%%%%%%%%%%%%%%%%%%%%%%%%%%%%%%%%
%%%%%%%%%%%%%%%%%%%%%%%%%%%%%%%%%%%%%%%%%%%%%%%%%%%%%%%%%%%%%%%%%%%%%
\pagebreak

\section{SPHERE+ Instrumental concept}% (5p) [Abo]
\label{sec:instru}

The main technical requirements necessary to achieve the science goals described in Section~\ref{sec:sciencecases} are: 1) to achieve higher contrasts at shorter separations, 2) to increase the sensitivity towards red targets, and 3) to perform better atmosphere characterization with medium to high spectral resolution. 

Improvement of the contrast close to the optical axis will come primarily from a 2$^{\rm nd}$-stage xAO system combined with a pyramid IR wavefront sensor, allowing faster correction (\texttt{tech.req.1}) and significantly better sensitivity to red targets (\texttt{tech.req.2}). Further improvement of contrast at short angular separations is provided by a combination of optimized coronagraphs, and algorithms for compensating non-common path aberrations. Some polarimetric components will be refurbished or modified to improve efficiency. SPHERE+ will also enable a full exploitation of the spectral dimension in high contrast imaging by providing higher spectral resolution than is currently possible, either in a narrow field with medium resolution, or in a single location of the field but at high resolution (\texttt{tech.req.3}). 
In the visible, a potential means to improve the contrast performance, with a focus on accretion signatures, involves a fast camera combined with an IFU.

\subsection{Adaptive Optics}

We propose an upgrade to the AO correction capability of SPHERE 
to provide improved raw contrast (deeper, closer) with {\bf a faster correction}, and which will enable observations of {\bf fainter (red) stars with an IR-PyrWFS}. 
Important gains will be secured through mature concepts and technological components that are already available, while on-going active research may provide %to potential
a significant boost to the baseline design. 
We choose {\bf a 2$^{\rm{nd}}$ stage AO correction approach} which involves a thorough development and test phase in Europe before implementation, and will allow the choice of the optimal design for fast measurement and correction in the small phase regime. 
SAXO's excellent AO performance and stability makes SPHERE the best facility worldwide for implementing our proposed upgrade.
\\

\noindent {\bf Second stage AO correction in the NIR}
\smallskip

As a second stage to be implemented in the current AO path, the new module takes short wavelength photons in the SAXO-corrected NIR path, and provides an additional faster-finer correction to the science instruments. The second stage includes its own specialized DM (with high speed but relaxed stroke requirements), sensor, and fast control law. This consistency with the existing AO system allows for {\bf stand alone development and test in Europe}, fed with well known SAXO correction residuals,  before implementation on SPHERE. The Phase A study will address in particular the questions of:

\begin{list} {$\bullet$}{\itemsep=0.1cm \parsep=-0.1cm \listparindent=0.0cm
   \labelsep=0.25cm \leftmargin=1.0cm \topsep=0.0cm}
    \item  interaction in between the 1$^{\rm{st}}$ stage and 2$^{\rm{nd}}$ stage AO loops. While this level of interaction is minimized in the proposed baseline - and such effect has already been addressed by our team - a detailed AO study will trade-off the control possibilities based on optimal algorithms, real time implementation constraints and performance requirements. 
   
   \item  2$^{\rm{nd}}$ stage RTC possibilities. Here a solution based on the COSMIC platform, already considered for other ESO instruments (MICADO, MAVIS) could fulfill this need. Other platforms such as SPARTA light could also be considered. In terms of implementation, two options are being contemplated as a starting point: either leveraging the existing SPARTA system for the first stage coupled to a dedicated RTC system for the second stage; or providing a new integrated solution driving both stages.  The obsolescence of SPARTA components will have to be addressed in collaboration with ESO, should the final concept depend on this platform. In any case, the considered data flow remains within reach of existing solutions and no technological show-stopper has been identified so far. Our team includes expertise in various design options, and is already engaged with other teams in Europe and Australia which could be leveraged to build a comprehensive development plan during phase A.
   
\item  opto-mechanical implementation within the current Common Path Instrument design in close relation between the AO correction study (LESIA) %, ONERA, LAM) 
    and CPI design (IPAG).
\end{list}

\vspace{0.1cm}
\noindent {\bf More Sensitive IR-PyrWFS}
\begin{figure}[h]
\begin{minipage}[c]{9cm}
    %\centering 
    \includegraphics[width=8.5cm]{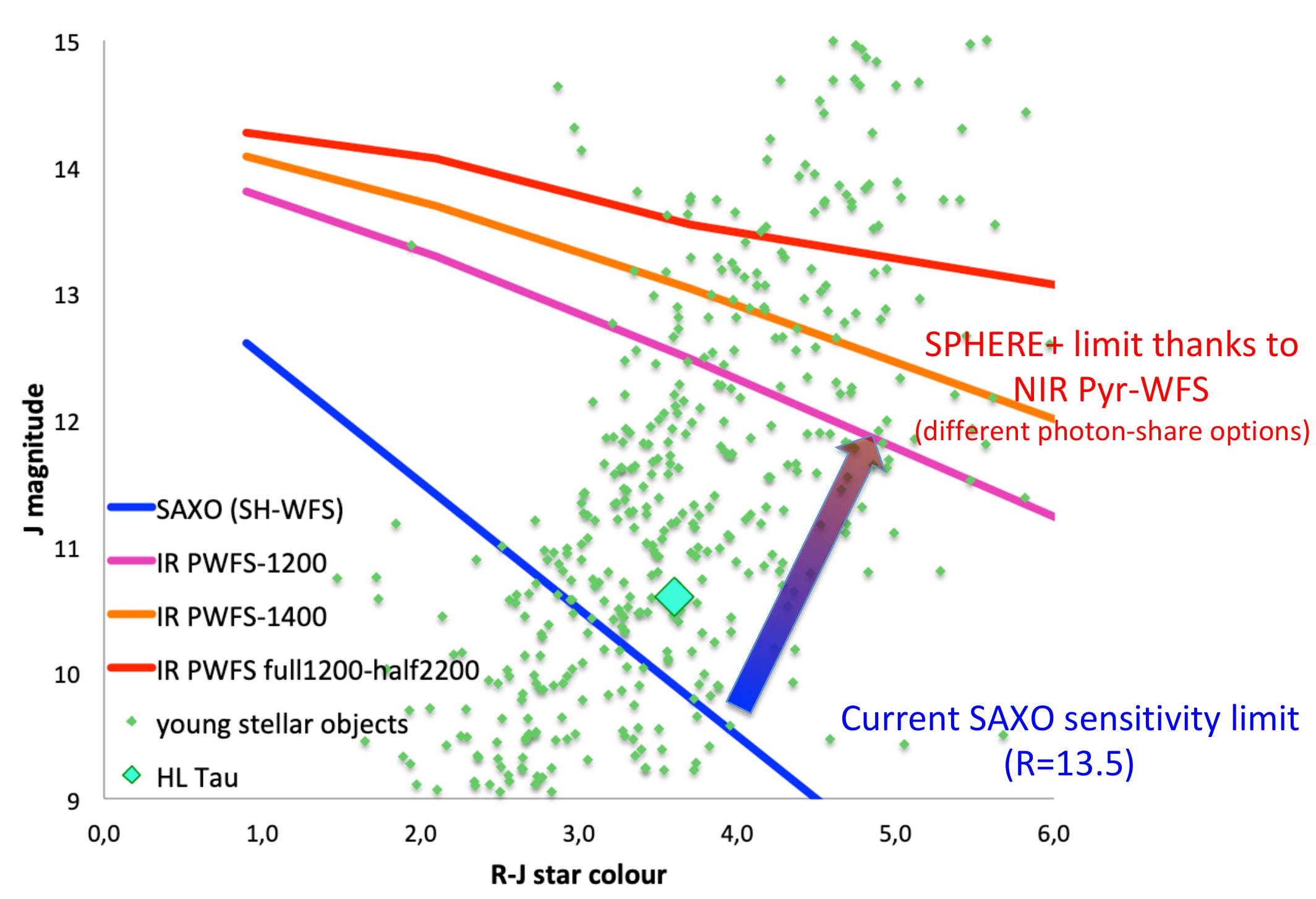}
\end{minipage}
 \begin{minipage}[c]{8cm}
   \caption{AO sensitivity limit gain thanks to an IR Pyramid sensor, as a function of the $R-J$ target colour, for different assumptions of photon share (all NIR photons to the sensor up to 1.2, 1.4\,$\mu$m or a 50/50 photon share in the 1.2 -2.2\,$\mu$m range). This gain will enable access to many more faint, red Young Stellar Objects: green dots represent actual targets from nearby star forming regions, among which the well-known example of HL\,Tau (currently out of reach of SPHERE).}
    \label{fig:saxo-sensitivity}
\end{minipage}
\end{figure}

The use of a pyramid wave-front sensor (PWFS) is an obvious choice in terms of improving pure wavefront sensing sensitivity. Invented 24 years ago, the PWFS concept has recently gained in expertise and maturity, mostly because of its implementation on all the first light AO modules of the ELTs, and on-going in-lab or on-sky demonstrator developments, in particular in our group (LESIA, ESO HOT experiment). % ,  LAM-ONERA,coll. INAF, Keck…).
The use of the near infra-red bands for wavefront sensing is also an attractive option motivated by the recent emergence of wide-band, extremely fast, sub-electron readout noise IR arrays. On top of its scientific interest for sensitivity on red faint targets (Figure \ref{fig:saxo-sensitivity}), wavefront sensing in NIR and behind SAXO allows a full-benefit of such “full-pupil” sensor, even more performing in small phase regime. The Phase A study will in particular address:

\begin{list} {$\bullet$}{\itemsep=0.1cm \parsep=-0.1cm \listparindent=0.0cm
   \labelsep=0.25cm \leftmargin=1.0cm \topsep=0.0cm}
\item     the optimal photon-share trade-off in terms of AO/science instrument sensitivity in various science cases (science group, instrument scientist, AO study). 
\item     the design optimization of the sensor exploring the specific case of operation without modulation (behind SAXO), and specific options (relation to on-going research program “WOLF” on optimal sensors).
\item     the integration of full feedback from on-going ELT studies, lab experiments and on-sky demonstrators (CANARDO experiment on WHT).%, coll Arcetri, Keck))
\item     the level of NCPA (in relation to the coronagraphy and dark hole studies, and WFS tolerance).
\end{list}

\begin{figure}[h]
\begin{minipage}[c]{9cm}
    %\centering 
    \includegraphics[width=8.5cm]{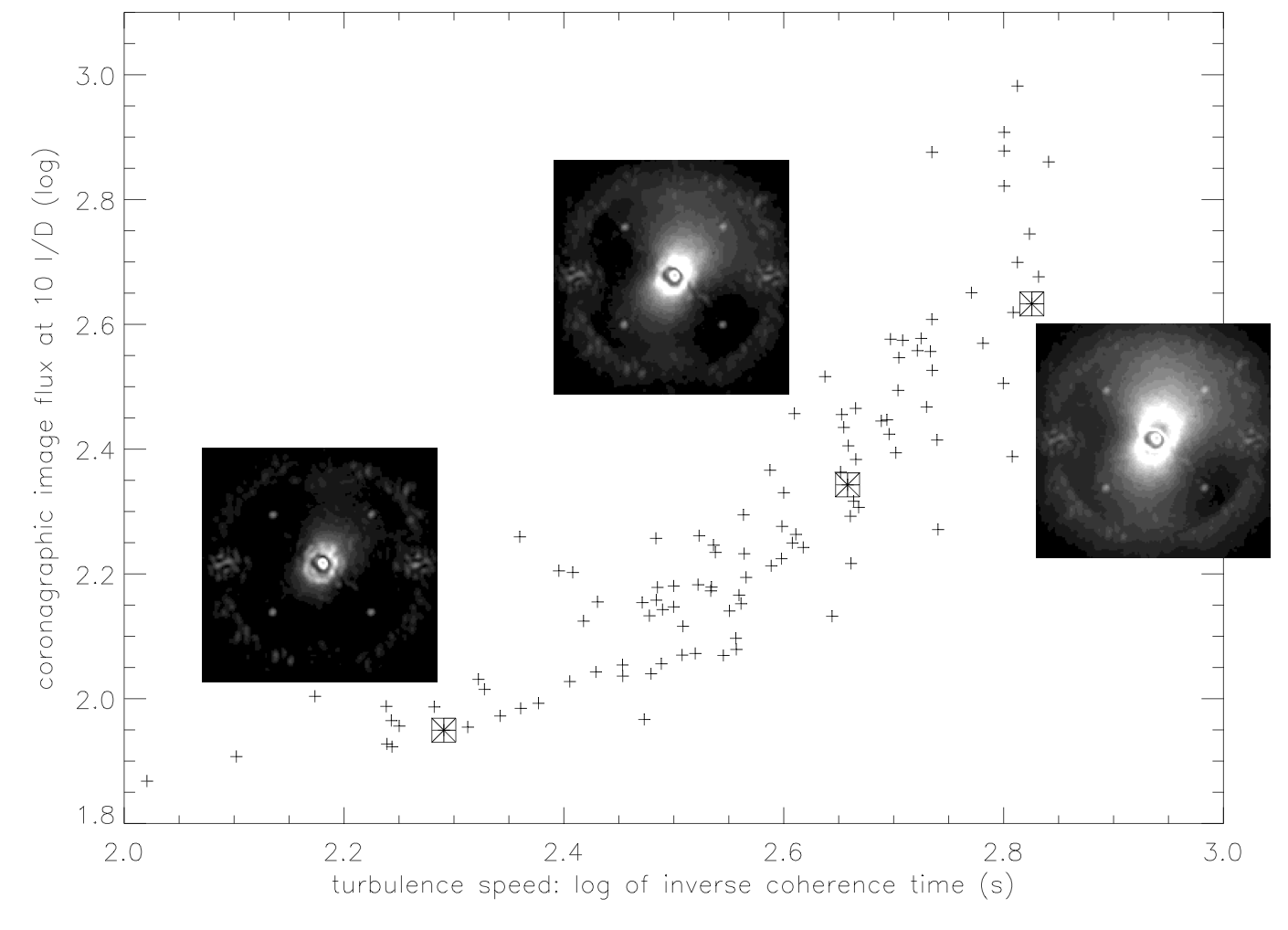}
\end{minipage}
 \begin{minipage}[c]{8cm}
     \includegraphics[width=8cm]{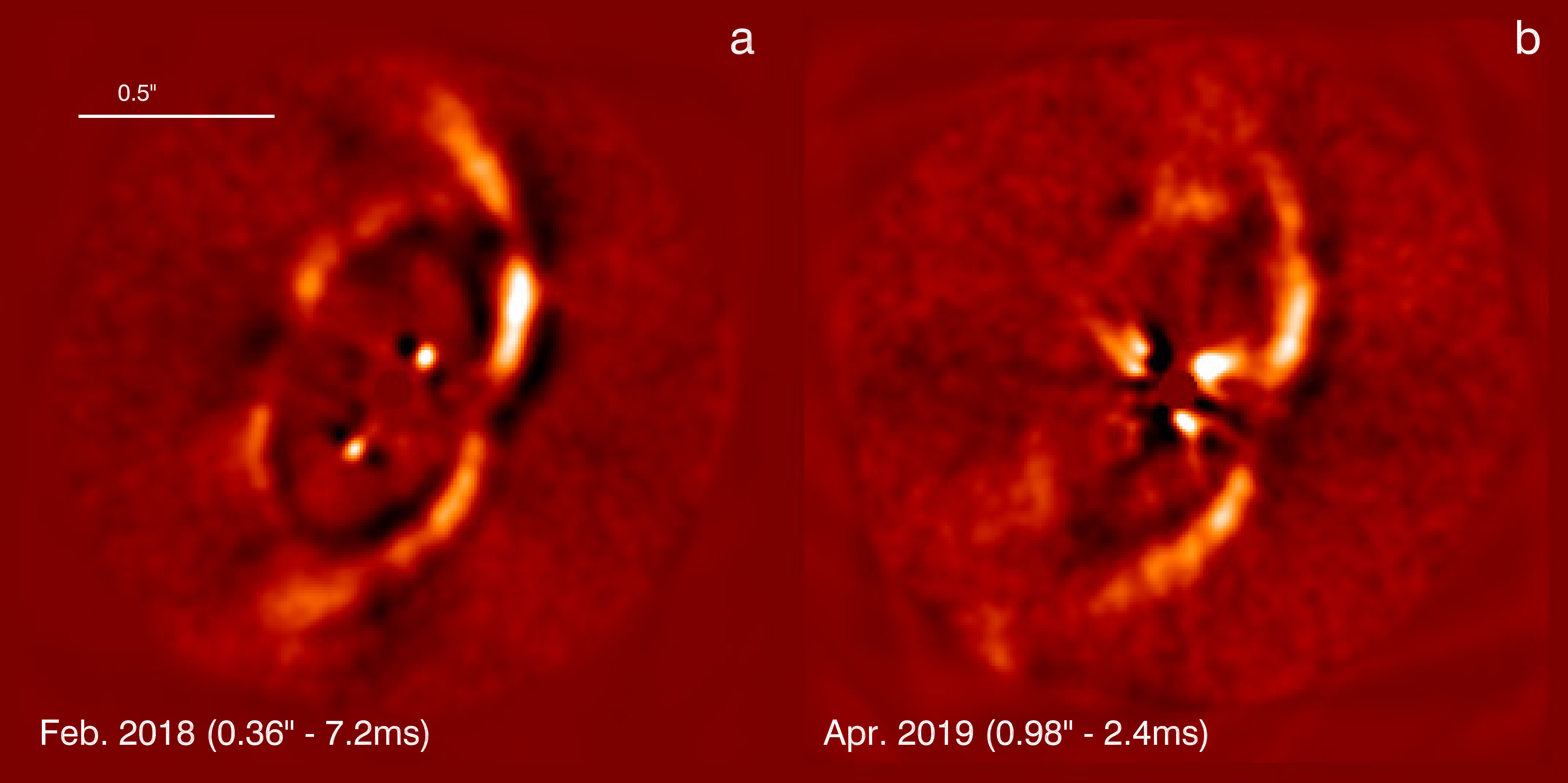}
  \caption{\textit{Left:} Correlation between the coronagraphic raw contrast at 10 $\lambda/D$ and the turbulence speed (inverse coherent time in log scale: slow turbulence on the left, faster towards right). \textit{Right:} PDS70 with the IFS at two epochs with a slow and fast turbulence. The planet is undetected in the second case.}
    \label{fig:saxo-perf-speed}
\end{minipage}
\end{figure}

\noindent {\bf Raw contrast and Faster Correction}
\smallskip

Years of SPHERE operations confirm the dominant role of AO servo-lag error on bright targets, with the raw contrast scaling as [AO correction / turbulence speed]$^{1.5}$ (Figure \ref{fig:saxo-perf-speed}, left). Pushing the frame rate to 3\,kHz would allow SPHERE+ to regularly attain the same level of results which are currently only obtained with SPHERE in the top 5\% observing conditions and would allow the highest contrast ratios to extend down to the 100 mas limit. Figure \ref{fig:saxo-perf-speed} (right) gives a sense of the gain for a famous object depending on the coherence time.
Technological components are currently available for this increased sampling frequency. Specific study will focus on optimizing the control law efficiency. This topic is an active field of research, with some preliminary tests already on sky.  Increased correction rate is mandatory for future high contrast imagers; SPHERE+  is clearly the best instrument for implementation of such an increased correction rate, and the concomitant increased science production that will result. 

\subsection{Near IR spectroscopy}

The science cases and associated TLRs require a higher spectral resolution than what SPHERE offers now; 
a resolution of a few thousands ($R_\lambda=5\,000-10\,000$) on one hand and a few tens of thousands ($R_\lambda=50\,000-100\,000$) on the other hand will dramatically improve its ability to characterize exoplanet atmospheres. 
While the current IFS could be modified to achieve a larger spectral resolution ($R_\lambda\approx1000$), a sensitivity analysis indicates that detector noise and thermal background become a serious limitation even if the IFS is moderately cooled down. The current IFS will still be needed for moderately wide FoV observations, so the Phase A will consider a medium-resolution IFU spectrograph (with a smaller FoV) and a high-resolution spectrograph for follow-up of known objects, both being distinct from the existing IFS.\\

\noindent {\bf HiRISE: High-spectral resolution in the near-infrared}
\smallskip

High spectral resolution with SPHERE can be achieved with the HiRISE concept proposed by \citet{Vigan2018}. HiRISE\footnote{High-resolution imaging and spectroscopy of exoplanets, \url{http://astro.vigan.fr/hirise.html}} will be a visitor instrument that implements a fiber connection between SPHERE and CRIRES+ to achieve $R_\lambda=100\,000$. The telecom fibers limit HiRISE to the $H$-band. 
SPHERE+ has the capability to enhance the science performance of  HiRISE. Therefore, SPHERE+ should be designed so as to improve the performance of HiRISE by increasing the throughput and providing access to the $K$-band with dedicated, high-transmission ZBLAN fibers.\\

\noindent {\bf MED-RES: medium-spectral resolution in the near-infrared}
\smallskip

A medium resolution integral field spectrograph will be a major addition to SPHERE and is needed to address the science requirements related primarily to exoplanets. 
We aim to achieve spectral resolution of at least $R_\lambda=5000$ consistent with the molecular mapping technique, and ideally $R_\lambda=10\,000$. 
The baseline design consists of a diffraction-limited IFU based on single-mode fibers or a single multi-core fiber covering the $J$ and $H$-bands where standard silica fibers are commercially available; extension to the $K$-band requiring special fibers will be examined during Phase A. The high fiber transmission allows the spectrograph to be placed at a remote location. The fiber head itself can be located on the back side of the platform being built for HiRISE, thereby minimizing the impact on SPHERE. 
%\textcolor{red}{Maybe write de sentence "extension to the K-band ... phase A" here after describing the JH case. RGa}
A fiber-fed spectrograph can also be combined with a dedicated apodizer that maximizes the coupling efficiency of off-axis point sources while suppressing the central star \citep{Por2018,Haffert2018}. Preliminary simulations show that a sensitivity gain of several magnitudes could be obtained in $J$ and $H$ very close to the star with such a system; this is to be weighted against possible throughput losses. A first prototype \citep{Haffert2018b} has been built and tested on-sky at the 4.2-m WHT in 2019. A detailed study during Phase A will identify the limitations of such a concept for SPHERE+. A comparison with alternative IFU schemes (lenslets/image slicers) will be performed. The IFU will feed one or several compact spectrographs with NIR detectors with low read-out noise and low dark current. 
Phase A will deliver a conceptual design of the MED-RES sub-system and an evaluation of its performance. Major trade-offs will include throughput, the number of spaxels, the spectral resolution and the (simultaneous) spectral range, the detectors and the noise associated with them, the number of spectrographs, compatibility with SPHERE, and the spectrograph design itself.

\subsection{Coronagraphy and non common path aberrations}

For an optimized high-contrast system, coronagraph designs must account for the AO architecture and performance, spectral bands, polarization effects, and their relation with non-common path aberrations (NCPA) compensation.  
While the current Apodized Pupil Lyot Coronagraphs (APLCs) are still considered efficient and will be offered, we have narrowed down the possible designs of additional coronagraphs by means of simulations of a modified APLC \citep{N'Diaye2015, N'Diaye2016b} and a Phase-apodized Pupil Lyot Coronagraph \citep{Por2020} which provide a contrast gain up to 10 in the 0.1-0.2$''$ range. %, which works well in tandem with the projected performance of an upgraded AO. 
In phase A we will confirm this finding with advanced simulations (especially regarding achromaticity) and investigate manufacturing aspects, implementation in SPHERE, and their relations to NCPA correction approaches and characterization modalities (spectroscopy and polarimetry). 

ZELDA is a phase-contrast technique that has already been tested on sky in SPHERE and which has proved efficient for correcting $\sim55$\,nm RMS NCPA residuals \citep{N'Diaye2016,Vigan2019}. 
For SPHERE+ we will explore solutions to actively operate ZELDA together with AO and with recent solutions in coronagraphic focal-plane wavefront sensing to provide further contrast gains \citep{Wilby2017,Bos2019}. 
In addition, we will produce a dark hole in the focal plane by using the combination of pair-wise probing (PW) and electric field conjugation (EFC).

\begin{figure}[!ht]
    \begin{minipage}[c]{0.6\textwidth}
    \includegraphics[width=0.41\textwidth]{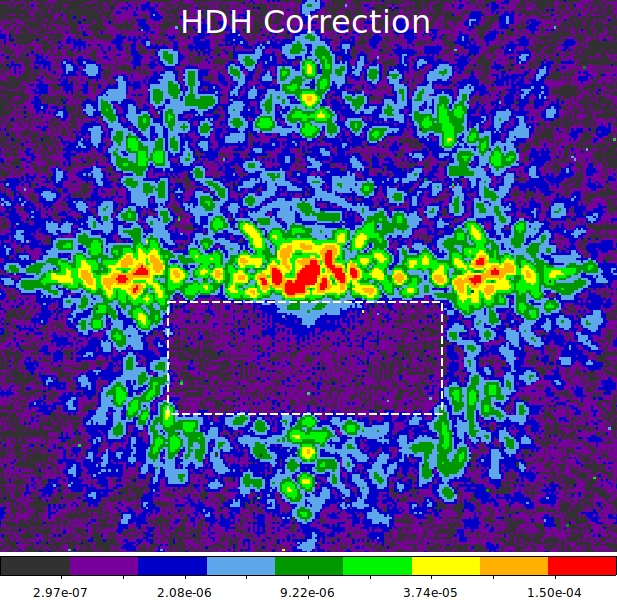}
    \includegraphics[width=0.58\textwidth]{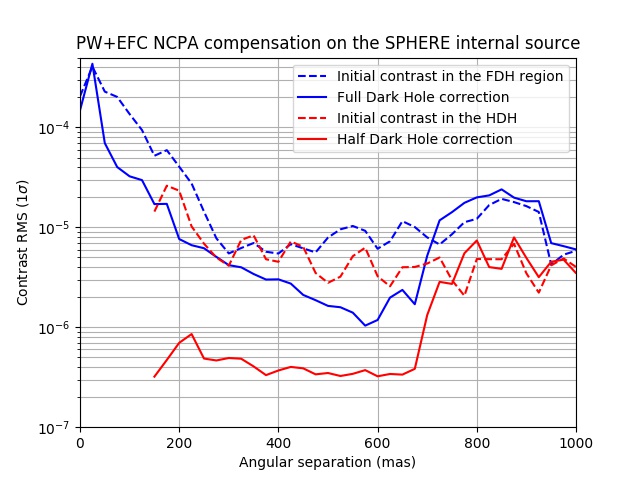}
  \end{minipage}\hfill
  \begin{minipage}[c]{0.39\textwidth}
    \caption{\textit{Left:} Coronagraphic science images after correcting NCPAs by PW+EFC on the SPHERE internal source in a half dark hole. \textit{Right:} azimuthal standard deviation before (dashed lines) and after (full lines) NCPAs compensation.}
    \label{fig:pwefc}
  \end{minipage}
\end{figure}

On SPHERE, PW-EFC reaches a contrast below $10^{-6}$ down to 0.2$''$ with the internal source (Potier et al. submitted, Figure\,\ref{fig:pwefc}). 
In phase A, we will investigate the on-sky performance, chromaticity, stability, and robustness of this technique that uses the AO to go beyond the starlight diffraction suppression provided by coronagraphs.

\subsection{Polarimetry}

While SPHERE polarimetry already produces spectacular results, 
we propose two improvements that can greatly boost the performance and expected science output of the polarimetric mode of IRDIS.
The current implementation of polarimetry for IRDIS leads to significant and variable instrumental polarization as a function of wavelength and pointing.
After rigorous calibration efforts by \citet{vanHolstein2020}, the instrumentally induced polarization and cross-talk can be corrected down to an absolute accuracy of $\sim$10$^{-3}$.
An issue that is not correctable in data reduction but well-understood is the fact that the efficiency of the polarization measurement is currently severely limited by the derotator coatings, and by the fact that the polarizing beam-splitter is implemented by two polarizers behind the non-polarizing beam-splitter of IRDIS. 
The latter implementation always throws away half of the light for polarimetry, whereas the derotator reduces the polarimetric efficiency down to $<$10\% at $H$ and $K$-bands \citep{vanHolstein2020}.
During the Phase A study for SPHERE+ we will study the replacement of the derotator and specifically its coatings to minimize its polarization effects while maximizing the reflectivity.
This study will also address the main cause of beam-shifts that are a major limitation for the polarimetric contrast performance of ZIMPOL \citep{Schmid2018}.
In addition, we will study the manufacturability of a high-quality, broadband polarizing beam-splitter for IRDIS.
These upgrades will enable IRDIS to perform more efficient polarimetric observations, and essentially add a standard polarimetric mode to any SPHERE observation to also detect circumstellar and/or circumplanetary scattering structures, whilst not relinquishing the current dual-band imaging capabilities.
The feasibility/risk of replacing the derotator and the IRDIS beam-splitter is an integral part of the study.
Moreover, we will investigate the implementation of new wave plates, to stabilize the instrumental polarization, and to add measurement capabilities for circular polarization.

\subsection{Visible channel}

The baseline AO configuration of SPHERE+ will not noticeably improve the wavefront correction in the visible channel, although phase A will revisit this assumption. \\

\noindent {\bf ZIMPOL:}

The upgrades foreseen in ZIMPOL are well identified and straighforward. In priority, better sensitivity, in particular in the  H$_\alpha$ line, is achievable with an additional detector mode providing low read-out noise, combined to a narrower filter. Also, off-axis observation will be implemented to reach separations as large as $4\,\!''$ from the central star.  
Finally, we will reassess the merits of all the masks of the visual coronagraph and the filters of ZIMPOL and we will consider new components in order to improve the science performance of the SPHERE visual channel \citep{Patapis2018}. 
As a prospective study we will explore the feasibility of polarimetric speckle mode,  which would allow diffraction limited resolution for strong circumstellar polarization signals.\\

\noindent {\bf IFS+FASTCAM for the Visible:}

Observations of accretion onto planets require high-contrast imaging in the visible at medium spectral resolution. Phase A will study the relevance and the feasibility of an additional sub-system to fully exploit the spatio-temporal and spectral information in the data. An IFS in the visible channel of SPHERE+ may gain at least one order of magnitude in contrast with respect to MUSE because of the better AO performance and spatial sampling. A high frame-rate and low readout-noise (EMCCD) detector would greatly help for faint targets and could freeze the atmospheric turbulence evolution to enable post-processing techniques similar to SHARK-VIS at the LBT \citep{LiCausi2017,Stangalini2017}. Phase A will study the coupling with SPHERE and consider different IFU options (fibers, lenslet array, image slicer) and detector options. Parameters to be considered are the field of view, pixel scale, fill factor, throughput, spectral resolution, spectral range, and mechanical limitations.

%%%%%%%%%%%%%%%%%%%%%%%%%%%%%%%%%%%%%%%%%%%%%%%%%%%%%%%%%%%%%%%%%%%%%%%
%%%%%%%%%%%%%%%%%%%%%%%%%%%%%%%%%%%%%%%%%%%%%%%%%%%%%%%%%%%%%%%%%%%%%%%
\pagebreak

\section{Conclusion}
SPHERE has been one of the highest-performing and most productive instruments in the last five years for high contrast imaging as confirmed by the number of refereed publications  (197 according to 
ADS). While exoplanetary science has been the main focus, the observing modes offered in SPHERE and its versatility has enabled an even broader range of science serving a large international community.  
SPHERE+ will build on this experience to push the main science cases, and will reveal a new population of young giant planets down to the snow line yet to characterize. At the same time, boosting the performances of SPHERE in terms of achievable contrast, sensitivity to redder stars, and spectral resolution will also foster new science cases that were marginally addressed with SPHERE. 

SPHERE+ operation must begin before 2026 in order to exploit the synergy with Gaia, RV surveys and ALMA given the competition, and to provide the most compelling targets for even more precise characterization: in particular with Gravity and Gravity+ for astrometry, and obviously for ELT instruments like MICADO, HARMONI and METIS, to search for and characterize the innermost planets in systems identified with SPHERE+. On the technical side, SPHERE+ is the ideal on-sky platform to test relevant components and strategies for the development of PCS (Planetary Camera and Spectrograph) at the ELT (second stage AO and high dispersion spectroscopy in particular).

To conclude, the SPHERE+ project with its proposed science cases and technical concept is certainly the most valuable instrument of its kind for furthering the field of exoplanet direct characterization, in strong synergy with several facilities like \textit{Gaia}, RV surveys, \textit{JWST}, today’s VLT/I instruments (including ESPRESSO, CRIRES+, ERIS, Gravity, Matisse) and the instrumentation road map of the ELT. 

%%%%%%%%%%%%%%%%%%%%%%%%%%%%%%%%%%%%%%%%%%%%%%%%%%%%%%%%%%%%%%%%%%%%%%%%
%%%%%%%%%%%%%%%%%%%%%%%%%%%%%%%%%%%%%%%%%%%%%%%%%%%%%%%%%%%%%%%%%%%%%%%%

%\pagebreak

%%%%%%%%%%%%%%%%%%%%%%%%%%%%%%%%%%%%%%%%%%%%%%%%%%%%%%%%%%%%%%%%%%%%%%%%%%
\pagebreak

\setlength{\bibsep}{0pt plus 0.3ex}

\bibliographystyle{aa} % style aa.bst
%{\footenotesize 
\bibliography{main}
%} % your references Yourfile.bib

\end{document}